\documentclass[12pt]{article}
\usepackage{comment}
\usepackage{cite}
\usepackage{qcircuit}
\usepackage{graphicx}
\usepackage[utf8]{inputenc}
\usepackage[dvips]{epsfig}
\usepackage{amssymb}
\usepackage[T1]{fontenc}
\usepackage{authblk}
\usepackage{amsmath}
\usepackage{float}
\usepackage{xcolor}
\usepackage{hyperref}
\usepackage{appendix}
\usepackage{subcaption}
\usepackage{multirow}
\usepackage[export]{adjustbox}
\usepackage{wrapfig}
\usepackage[english]{babel}

\title{\textbf{Entanglement detection with quantum support vector machine (QSVM) on near-term quantum devices}}

\author[1,2,3]{Mahmoud Mahdian\thanks{mahdian@tabrizu.ac.ir}}
\author[1]{Zahra Mousavi\thanks {z.mosavi1400@ms.tabrizu.ac.ir}}
\date{\today}

\affil[1]{Faculty of Physics, Theoretical and Astrophysics
	Department, University of Tabriz, 51665-163 Tabriz, Iran}
\affil[2]{Research Institute for Applied Physics and Astronomy (RIAPA), University of Tabriz, Tabriz, Iran}
\affil[3]{Quantum Technology Center, University of Tabriz, Tabriz, Iran}
\begin{document}
\maketitle
\begin{abstract}

Detecting and quantifying quantum entanglement remain significant challenges in the noisy intermediate-scale quantum (NISQ) era. This study presents the implementation of quantum support vector machines (QSVMs) on IBM quantum devices to identify and classify entangled states. By employing quantum variational circuits, the proposed framework achieves a runtime complexity of $O(\frac{N t}{\epsilon^2})$, where $N$ is the number of qubits, $t$ is the number of iterations, and $\epsilon$ is the acceptable error margin. We investigate various quantum circuits
with multiple blocks and obtain the accuracy of QSVM as measures of expressibility and entangling capability. Our results demonstrate that the QSVM framework achieves over 90\% accuracy in distinguishing entangled states, despite hardware noise such as decoherence and gate errors. Benchmarks across superconducting qubit platforms (e.g., IBM Perth, Lagos, and Nairobi) highlight the robustness of the model. Furthermore, the QSVM framework effectively classifies two-qubit states and extends its predictive capabilities to three-qubit entangled states. This work marks a significant advancement in quantum machine learning for entanglement detection.

\end{abstract}

\section{Introduction}

Entanglement is one of the most fascinating phenomena in quantum mechanics, representing the intricate quantum correlations between particles that persist regardless of spatial or temporal separation \cite{Nielsen, Einstein, Schrodinger}. This unique property is a cornerstone of quantum technologies, enabling breakthroughs in quantum computing \cite{shor, Nielsen}, quantum cryptography \cite{gisin2002g}, quantum communication \cite{Bennett1993}, and beyond. To fully harness the potential of entanglement, it is essential to develop robust methods for its detection and characterization.\\
However, detecting entanglement, particularly in the Noisy Intermediate-Scale Quantum (NISQ) era, presents significant challenges. Traditional methods, such as the positive partial transpose (PPT) criterion \cite{peres, Horodecki}, are effective for low-dimensional systems (e.g., $\mathcal{H}^2 \otimes \mathcal{H}^2$ and $\mathcal{H}^2 \otimes \mathcal{H}^3$) but fail to provide sufficient conditions for separability in higher dimensions. For instance, entangled states with positive partial transpose (PPTES) \cite{pHorodecki} cannot be identified using the PPT criterion alone. Another method to distinguish separable states from entangled ones involves the use of entanglement witnesses (EWs) \cite{Terhal,jafarizadeh,Mahdian,Jamiolkowski}, which are Hermitian operators. However, while EWs are a valuable tool, their practical application often demands precise construction and is constrained by the complexity of high-dimensional systems. These challenges are exacerbated in NISQ devices, where noise, decoherence, and limited qubit connectivity further complicate entanglement detection.

Machine learning (ML) is a subset of artificial intelligence (AI) that enables computers to learn from data without being explicitly programmed. Machine learning algorithms construct mathematical models from sample data, referred to as “training data,” to make predictions or decisions autonomously. These algorithms adapt to new data, producing reliable and repeatable results \cite{Mohri}. Support Vector Machines (SVMs) are a prominent type of supervised machine learning algorithm, well-regarded for their ability to classify data points effectively by identifying the optimal hyperplane that separates different classes. However, as datasets become larger and more complex, traditional SVMs encounter challenges related to scalability and computational efficiency. This is where quantum computing plays a crucial role. Authors in \cite{PhysRevApplied.19.034058} successfully employed SVM algorithm to obtain entangled witnesses, distinguishing entangled states from separable ones. Neural networks, another type of supervised machine learning algorithm, can also be utilized to detect quantum entanglement. Similar to SVMs, neural networks utilize labeled data to determine the best decision boundary between two classes \cite{Entanglementdetection}.

In recent years, the prospect of using quantum devices to help or completely replace by classical systems in machine learning has made great progress and combination of machine learning and quantum computing has the potential to revolutionize artificial intelligence \cite{Schuld,PhysRevX.8.021050,doi:10.1080/00107514.2014.964942,PhysRevResearch.1.033063}.
Quantum machine learning (QML) is an emerging subfield of artificial intelligence that leverages the principles and techniques of quantum computing to enhance the capabilities of machine learning algorithms in data analysis \cite{mainref, jager2023universal, kavitha2022quantum}.A ke By exploiting unique quantum advantages, QML aims to develop novel algorithms capable of processing data more efficiently and accurately, with the ultimate goal of creating more powerful and versatile AI systems. A key development in this field is the emergence of hybrid variational quantum-classical algorithms, which utilize low-depth quantum circuits to efficiently combine quantum and classical resources. These algorithms are particularly well-suited for QML, as they can address specific tasks beyond the reach of traditional classical computers while operating within a framework that imposes less stringent hardware requirements, making them promising for near-term applications \cite{Preskill, Peruzzo, Farhi, Otterbach, McClean, MahmoudMahdian, MahmoudMahdian11}. 

The study of quantum entanglement has significantly progressed with QML techniques for detecting and measuring entanglement by utilizing quantum states as inputs. QSVMs represent a promising intersection of quantum computing and machine learning, offering potential advantages over classical SVMs by leveraging quantum algorithms to solve complex classification problems more efficiently. QSVMs utilize quantum principles, such as superposition and entanglement, to process high-dimensional data and solve optimization tasks that are computationally expensive for classical systems. For instance, Rebentrost et al. \cite{PhysRevLett.113.130503} introduced QSVMs for binary classification, demonstrating their ability to achieve exponential speedups in certain scenarios by using quantum algorithms like the Harrow-Hassidim-Lloyd (HHL) algorithm for solving linear systems \cite{PhysRevLett.103.150502}. In quantum computing, QSVMs are being explored for applications such as quantum state classification \cite{PhysRevA.97.042315}, quantum chemistry \cite{sajjan2022quantum}, and quantum finance \cite{ORUS2019100028}, where they can classify molecular structures, optimize financial portfolios, and detect anomalies in quantum systems. Additionally, QSVMs are being investigated for quantum image processing\cite{10.21203/rs.3.rs-1434074/v1} and predicting air pollution \cite{farooq2024enhanced}, highlighting their versatility across domains. Despite their potential, challenges such as error rates in quantum hardware and scalability remain, necessitating further research to fully realize their capabilities.

In this study, we introduce a novel QSVM classifier utilizing variational quantum circuits (VQCs) tailored for NISQ devices \cite{nisq-era, Variational-quantum-algorithms}. VQCs, which consist of parameterized quantum gates, are central to variational quantum algorithms (VQAs) and are optimized imperatively using classical methods. We
investigate the expressibility and entanglement capability of various VQCs to enhance the performance and reliability of QSVMs. Our approach leverages quantum circuits to prepare training data that reflects the quantum characteristics of the dataset, enabling the detection of entangled and separable states with high accuracy. Notably, our method achieves over $\%90$ accuracy in predicting two and three-qubit states, demonstrating its effectiveness even with limited training data.\\
The contributions of this work are threefold: a scalable entanglement detection framework designed for NISQ devices, a comprehensive evaluation of VQC properties (expressibility and entanglement capability) in the context of QSVMs, and experimental validation on IBMQ devices, showcasing
the practical applicability of our approach. By addressing the limitations of existing methods and leveraging the strengths of QML, this study paves the way for more efficient and reliable entanglement detection in quantum
systems.

The remainder of this paper is organized as follows: Section II presents a framework for general QSVMs and the preparation of two- and three-qubit entangled and separable quantum states using quantum circuits. In Section III, we discuss the results obtained from experiments conducted on IBMQ. Finally, we conclude the paper and suggest avenues for future research.

\section{Quantum Support Vector Machines}

Quantum states are characterized by positive trace-class operators, which are Hermitian operators acting on a Hilbert space. These states, known as density matrices $\rho$, are defined as:

$$\mathcal{S}(\mathcal{H}) \equiv \left\{ \rho \in \mathcal{T}(H) \, \middle| \, \rho \geq 0, \, \text{tr}(\rho) = 1 \right\}.$$
Quantum states can be categorized as separable or entangled. Entanglement is a phenomenon where two or more quantum systems are correlated in such a way that the state of one system cannot be described independently of the others. The Hilbert space of a two-level quantum system, or qubit, is represented by the Bloch sphere. The algebra of observables for this space is derived from the Lie group SU(2), generated by the Pauli matrices. A multi-qubit system, defined on the Hilbert space$\mathcal{H}_{qubit}^{\otimes N}=\mathcal{H}_{2}^{(1)}\otimes\mathcal{H}_{2}^{(2)}\otimes...\otimes\mathcal{H}_{2}^{(N)}$ 
is called separable if it can be expressed as:

\begin{equation}\label{eq:state}
\mathcal{\rho} = \sum_{i=1} p_i \rho^{(1)}\otimes...\otimes\rho^{(N)},\ \ p_i\geq 0  \ \ \sum _i p_i=1,
\end{equation}

where $\rho^{(i)}$
are arbitrary, normalized single-qubit states in $\mathcal{H}_{2}^{(i)}=\mathcal{C}^2$. If $\rho$ cannot be written in this form, it is considered an entangled state.

In QSVM algorithms, two main classification methods are used: the kernel trick method and the variational quantum circuit (VQC) method. This study focuses on classifying entangled and separable states using the VQC-SVM approach. The dataset is structured as $\{\zeta\equiv\{({\rho_m}, y^m)\}_{m=0}^M, where\ \rho_m\in \mathcal{H}_{qubit}^{\otimes N}, y^m = \pm 1 \}$. For binary classification, the label $(-1)$ represents entangled states, while $(+1)$
represents separable states.

The optimization problem for the SVM is formulated in its dual form, applying the Karush-Kuhn-Tucker (KKT) conditions, leading to the following loss function:

\begin{equation} \label{eq:soft dual la}
\mathcal{L}(\mu, C,\lambda) = \frac {1}{C} \sum_{i=1}^{M} \mu_i^2 + \frac{1}{2}\sum_{i,j=1}^{M} \{y_i y_j \mu_i \mu_j (\rho_i^T \rho_j) - \frac{1}{2\lambda}\} ,
\end{equation}

where $C$ is the regularization parameter that controls the trade-off between minimizing training error and model complexity, and $\lambda$ penalizes large values of $\mu$, thereby preventing overfitting (for more details see Appendix A).

For a new state $\boldsymbol{\varrho}$ the decision function for classification is given by :
\begin{equation}\label{eq:decision soft}
\begin{aligned}
{f}(\boldsymbol{\varrho}) = \text{sgn}\left(\sum_{i=1}^{M} \mu_i y_i \boldsymbol{\rho}_i^T \boldsymbol{\varrho} + \frac{1}{\lambda}\right),
\end{aligned}
\end{equation}

where the sign function \text{sgn}(·) is defined as:

\[
\text{{sgn}}(\varrho) =
\begin{cases}
    -1, & \text{{if }} \varrho\  is  \ entangled ; \\
    1, & \text{{if }} \varrho \  is  \ seperable.
\end{cases}
\]
\\ 
In this work, we convert the classical SVM into a QSVM by employing variational quantum circuits (VQCs). These circuits are parameterized quantum circuits designed to optimize the Lagrange coefficients $\mu$, which depend on the rotation angles of individual qubits. A PQC (Parameterized Quantum Circuit) is used to generate parametric states:

\begin{equation}\label{eq:quantum dual_1}
|\psi (\boldsymbol{\vec{\theta}})\rangle=U\left(\boldsymbol{\vec{\theta}} \right) | 0\rangle ^{\otimes N} ,
\end{equation}

where the unitary operator $U\left(\boldsymbol{\vec{\theta}} \right)$
, also known as the ansatz, acts on the reference state. By adjusting the parameters 
$\vec{\theta}$, the final quantum state is controlled. Depending on the problem, different types of ansatzes, such as the unitary coupled-cluster, low-depth circuit ansatz, or SWAP network, can be employed \cite{Romero_2019,Dallaire-Demers_2019,PhysRevA.106.042433,PhysRevLett.120.110501}.

\begin{figure}
	\centering
	\begin{subfigure}[H]{0.4\textwidth}
		\centering
		\includegraphics[width=\textwidth]{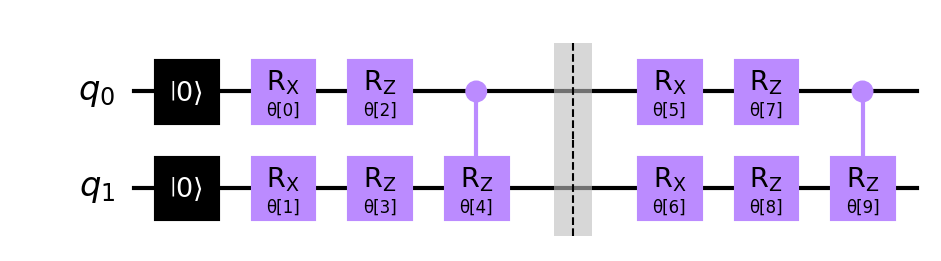}
		\caption{}
		\label{fig:var qubit (1)}
	\end{subfigure}
	\hfill
	\begin{subfigure}[H]{0.4\textwidth}
		\centering
		\includegraphics[width=\textwidth]{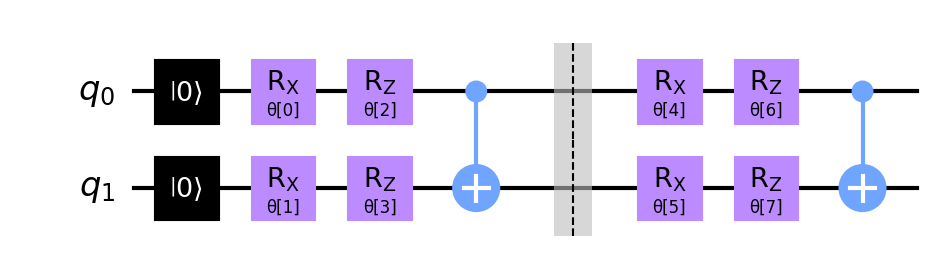}
		\caption{}
		\label{fig:var qubit (2)}
	\end{subfigure}
	\hfill
	\begin{subfigure}[H]{0.4\textwidth}
		\centering
		\includegraphics[width=\textwidth]{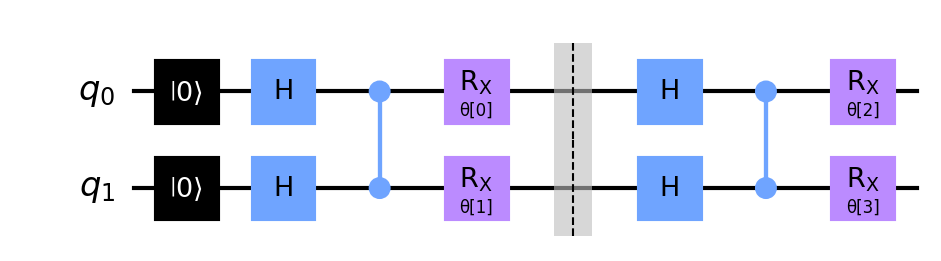}
		\caption{}
		\label{fig:var qubit (3)}
	\end{subfigure}
	\caption{Architectural design of two-qubit VQC circuits. (1)  (\# circuit 1). (2) (\# circuit 2). (3) (\# circuit 3).}
	\label{fig:vqcc}
\end{figure}

\begin{figure}
	\centering
	\begin{subfigure}[H]{0.4\textwidth}
		\centering
		\includegraphics[width=\textwidth]{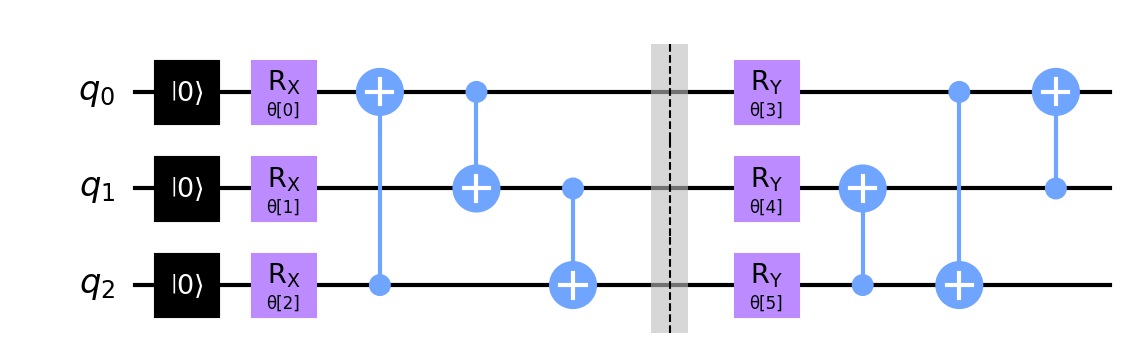}
		\caption{}
		\label{fig:var qubitt (4)}
	\end{subfigure}
	\hfill
	\begin{subfigure}[H]{0.4\textwidth}
		\centering
		\includegraphics[width=\textwidth]{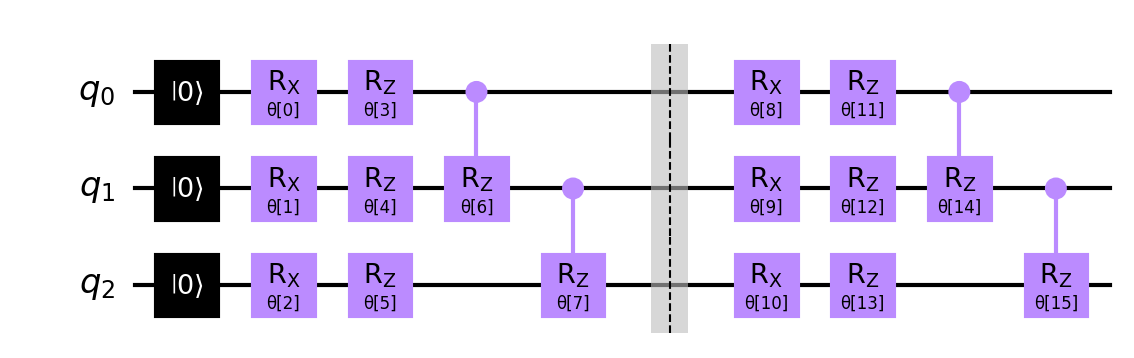}
		\caption{}
		\label{fig:var qubitt (2)}
	\end{subfigure}
	\hfill
	\begin{subfigure}[H]{0.4\textwidth}
		\centering
		\includegraphics[width=\textwidth]{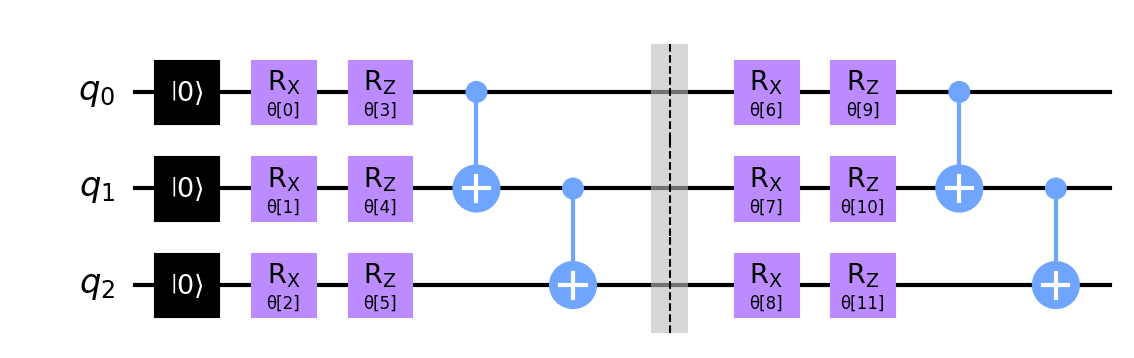}
		\caption{}
		\label{fig:var qubitt (1)}
	\end{subfigure}
	\hfill
	\begin{subfigure}[H]{0.4\textwidth}
		\centering
		\includegraphics[width=\textwidth]{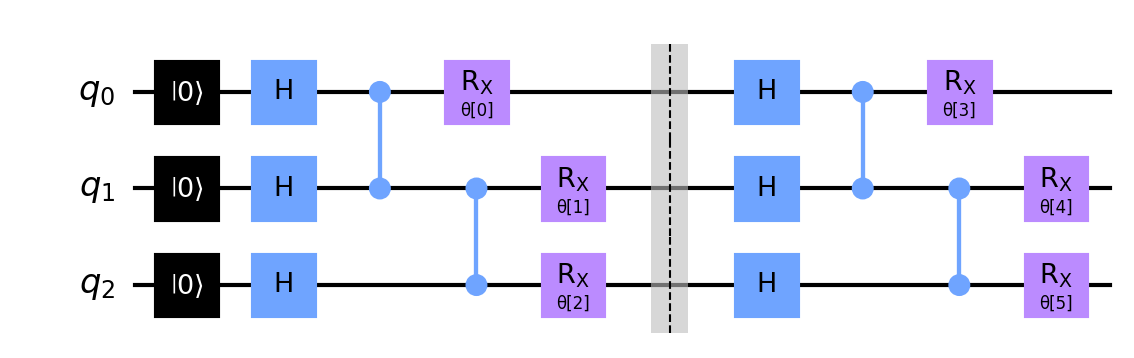}
		\caption{}
		\label{fig:var qubitt (3)}
	\end{subfigure}
	\caption{Architectural design of three-qubit VQC circuits. (1)  (\# circuit 1). (2) (\# circuit 2). (3) (\# circuit 3). (4) (\# circuit 4).}
	\label{fig:vqc}
\end{figure}

In this paper, we use a PQC that can be implemented on near-term hardware, which consists of single-qubit gate layers with adjustable parameters, as well as two-qubit gate layers to create entanglement as shown below:

\begin{equation}\label{eq:quantum dual_2}
U\left(\boldsymbol{\vec{\theta}} \right)=\prod_k U (\theta_k) W_k,
\end{equation}

\begin{equation}\label{eq:quantum dual_3}
U (\theta_k) =R( \theta_k ^i)^{\otimes n}= \bigotimes _n e^{-i \theta_k ^i \sigma /2} ,
\end{equation}

where  $\sigma$ represents one of the Pauli matrices, and  $W_k$ represents a two-qubit gate. The parameters 
$\boldsymbol{\vec{\theta}}=(\theta_1,\theta_2,...)$
determine the transitions between states, enabling the PQC to represent the classical decision function in quantum form.
The loss function depends on the VQC technique (ansatz) and consists of a series of quantum gates and operations ( see Figuers of .\ref{fig:vqcc} and \ref{fig:vqc} ). The optimization parameters $\mu_i (\boldsymbol{\vec{\theta}})$ are derived from the ansatz and are equivalent to the squared magnitude of the expectation value of the operator 
$U(\boldsymbol{\vec{\theta}})$ applied to the initial state.
 As outlined in Appendix A Eq.\ref{eq:soft dual laa}, we adopt the classical SVM loss function method, enhanced by the integration of quantum-based parameters for improved performance.The quantum-based loss function is expressed as:
\begin{equation}\label{eq:quantum dual_4}
	\mathcal{L}_{\psi, \gamma,C}(\boldsymbol{\vec{\theta}} , \zeta)=\sum_{i=1}^{M} \sum_{j=1}^{M} \mu_i(\boldsymbol{\vec{\theta}}) \mu_j(\boldsymbol{\vec{\theta}}) y_i y_j\left[\left|\left\langle\psi_i \mid \psi_j\right\rangle\right|^2+\frac{1}{\gamma}\right]+ \frac{1}{C} \sum_{i=1}^{M} \mu_i(\boldsymbol{\vec{\theta}})^2.
\end{equation}

\begin{figure}[H]
	\centering
	\includegraphics[width=0.8\linewidth, height=5.5cm]{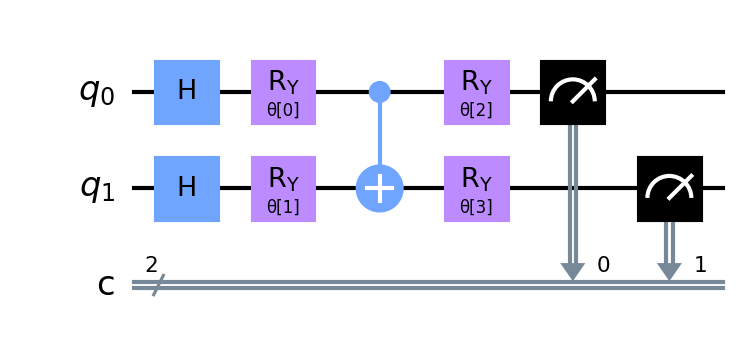}
	\caption{Two-qubit $\mu_i (\boldsymbol{\vec{\theta}})$'s Circuit using four rotation gates and a CNOT. }
	\label{fig:ALPHA I}
\end{figure}

This framework allows for the efficient detection of entangled and separable quantum states, paving the way for broader applications of QSVMs in quantum machine learning(see Fig. ~\ref{fig:ALPHA I} ).

\begin{figure}[H]
	\centering
	\includegraphics[width=0.8\linewidth, height=5cm]{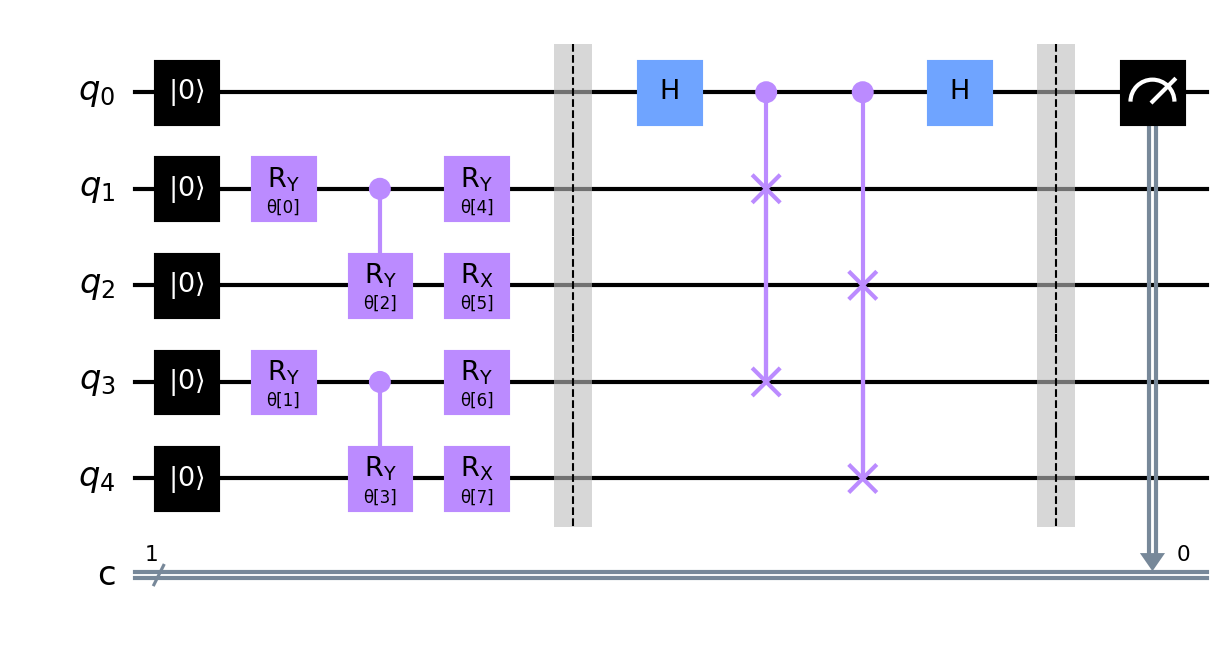}
	\caption{The swap gate operation for to calculate dot product states.}
	\label{fig:swap}
\end{figure}

The measurement results of all qubits provide the probabilities for each basis. The lost function of eq. \eqref{eq:quantum dual_4} is mostly based on the evaluation of the dot product which can be estimated via the Swap test subroutine ~\cite{beyondkernelmethods,featureHilbertspaces,swap1,swap2}.
Figure \ref{fig:swap} performs the inner product between two quantum state vectors and upon measuring the ancilla qubit, we obtain the count number in two bases $|0\rangle$ and $|1\rangle$. Finally, the expression $R = 1 - \frac{2}{S} \sum_{i=1}^{S} M_{i}$ describes a calculation where the result $R$ is influenced by the number of shots $S$ and $M$ (number of counts for the basis  $|1\rangle$)~\cite{Quantumfingerprinting,swap1,nearest_neighbor}.

\begin{figure}[H]
	\centering
	\includegraphics[width=0.75\linewidth, height=5.7cm]{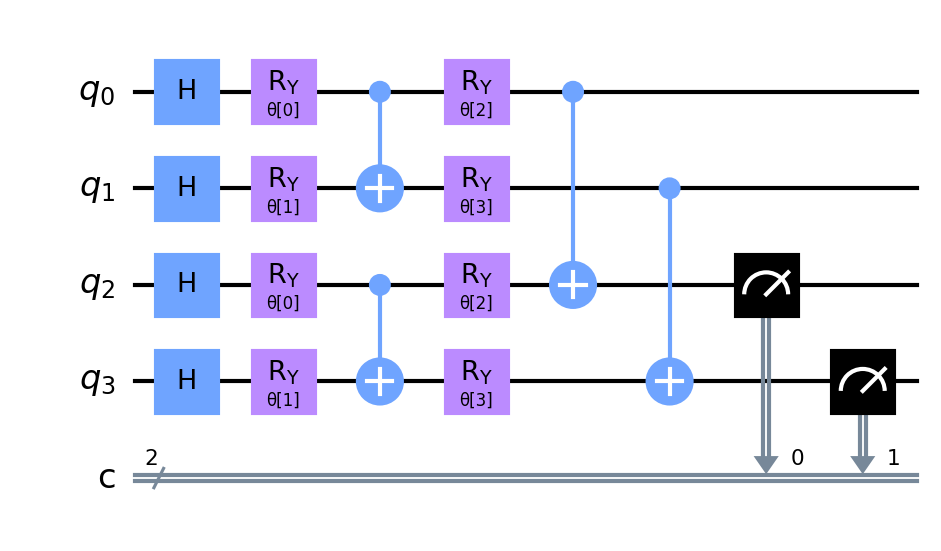}
	\caption{The final goal of the second factor in the Eq.~\eqref{eq:quantum dual_4} was prepared with measurement results on $M_{00..0}$ where $M$  is a projection measurement operator of state and disparity will be greater for smaller shot sizes.}
	\label{fig:r_theta}
\end{figure}

\renewcommand{\thesubfigure}{\arabic{subfigure}}

After setting up the objective function, we need to minimize it in order to find optimal values of $\boldsymbol{\vec{\theta}}$.
The decision function is represented in Eq.\eqref{eq:quantum dual op} and in this stage, optimal $\boldsymbol{\vec{\theta}}^*$s are provided as input. Finally the predicted test data labels reached from $\hat{y}=\text{{sgn}}(f_{\psi, \gamma}(\boldsymbol{\boldsymbol{\vec{\theta}}^*} , \zeta,\psi_{test})) $.
\begin{equation}\label{eq:quantum dual op}
f_{\psi, \gamma}(\boldsymbol{\boldsymbol{\vec{\theta}}^*} , \boldsymbol{\zeta},\psi_{test})=\sum_{i=1}^{M} \mu_i(\boldsymbol{\boldsymbol{\vec{\theta}}^*}) y_i \left[\left|\left\langle\psi_i \mid \psi_{test}\right\rangle\right|^2+\frac{1}{\gamma}\right].
\end{equation}

Alternatively, two-class registers are defined as $i$ $(q_0 ,\ q_1)$ and $j$ $(q_2 ,\ q_3)$ in Fig~\ref{fig:r_theta}. Utilizing the ansatz $U(\boldsymbol{\vec{\theta}})$ for $N=4$, the initial two qubits are assigned to register class one, while the subsequent two qubits are designated for another class. 

\subsection{Data Preparation and Feature Sampling }

The primary goal of this article is to determine whether a quantum states is entangled or separable using QSVM. So, the starting point is to collect data on quantum states, in two categories. Here we discuss two-qubit, three-qubit, and bipartite states.
\subsubsection{Sepearble states}
Using \ref{eq:state}, we can construct completely separable mixed states and also, any pure state on the Bloch sphere can be obtained according to the rotation as follows:

\begin{equation}\label{eq:phi1}
\begin{aligned}
   \sigma_{sep} &= \sum_i p_i |\varphi(\alpha_i,\beta_i)\rangle \langle\varphi(\alpha_i,\beta_i)\rangle ,\\
    \end{aligned}
\end{equation}

where
\begin{equation}\label{eq:phi22}
\begin{aligned}
|\varphi(\alpha_i,\beta_i)\rangle=\bigotimes_j^N \cos(\alpha_{ij})|0_j\rangle +e^{\beta_{ij}} \sin(\theta_{ij})|1_j\rangle=\bigotimes_j^N U(\alpha_{ij},\beta_{ij}) |0_j\rangle,
\end{aligned}
\end{equation}

with
\begin{equation*}
U(\alpha_{ij},\beta_{ij}) =
\begin{pmatrix}
\cos(\alpha_{ij}) &-\sin(\alpha_{ij}) \\
\sin(\alpha_{ij}) & e^{\beta_{ij}}\cos(\alpha_{ij})
\end{pmatrix},
\end{equation*}
which can be considered as rotation around axis y, $R_y(\alpha_{ij})$, and then around axis z,$R_z(\beta_{ij})$. So,it is possible to create a separable quantum state by operating single-qubit gates on NISQ device.
For a two-qubit quantum state, the parameter 
	$|\varphi\rangle$ can be explained as follows:

\begin{equation}
		|\varphi(\alpha_i,\beta_i)\rangle=\bigotimes_{j=1}^2 U(\alpha_{i1},\beta_{i1}) |0_j\rangle \otimes U(\alpha_{i2},\beta_{i2}) |0_j\rangle, 
\end{equation}
and unitary transformations operate locally, affecting individual qubits independently.

\subsubsection{Entangled states}
\begin{figure}[H]
	\centering
	\includegraphics[width=0.53\linewidth, height=4cm]{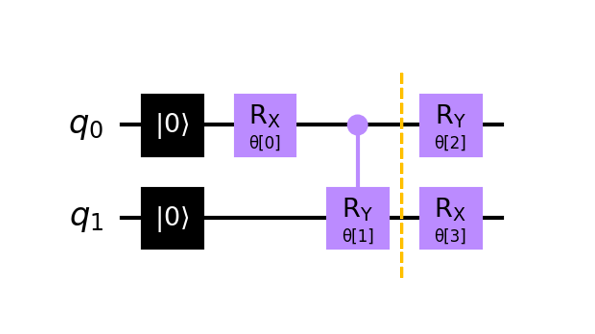}
	\caption{Two-qubit state preparation circuit. The  $|\psi \rangle$ extracted from the highlighted line is fundamental to the computation of the concurrence method. The state is $|\psi \rangle = \cos(\frac{\theta_0}{2}) |00 \rangle - i \sin(\frac{\theta_0}{2}) \cos(\frac{\theta_1}{2}) |10 \rangle -
	 i \sin(\frac{\theta_0}{2}) \sin(\frac{\theta_1}{2}) |11 \rangle $. For $\theta_0 = \frac{\pi}{2} ,\theta_1 = \pi ,  $ the state becomes maximally entangled (c=1),  while for  $\theta_0 = 0 ,\theta_1 = \pi ,$ the state is separable (c=0). } 
	\label{fig:2 qubit}
\end{figure}

A circuit-based data for creating a entangled two and three qubits states are demonstrated in~\cite{Two-qubitBlochsphere,ghz}.\\ 
\textit{Two-qubit}:The concurrence method for preparing a range of minimum and maximum entangled states in terms of the rotation angles of the unitary gates. Let $c$ be a parameter defined as follows:

\begin{align*}
\text{Concurrence} =
\begin{cases}
    c = 0 & \quad \text{ separable states}; \\
    0 < c < 1 & \quad \text{ entangled states}; \\
    c = 1 & \quad \text{maximally entangled states}.
\end{cases}
\end{align*}
Concurrence can be described as an informative item in quantum circuits for labeling data and its value in this circuit is equivalent to $c=\sin(\theta_0) \sin\left(\frac{\theta_1}{2}\right)$ where $\sin(\theta_0)$ is the control qubits degree of initial superposition (see Fig.~\ref{fig:2 qubit}).

\textit{Three-qubit}: The GHZ-class can be driven from Eq.\eqref{eq:ghz}. The five parameter's respective ranges are $\theta_{A,B,C} \in \left(0, \pi/2\right] $, $\vartheta \in \left(0, \pi/4\right]$,  $\phi \in \left[0, 2 \pi\right)$, $P\in \left(\frac{1}{2}, \infty\right)$. Here, $P$ represents a normalization factor.

\begin{equation}\label{eq:ghz}
    |\psi_{\text{GHZ}}\rangle = \frac{1}{\sqrt{P}}\left(\cos(\vartheta)|000\rangle + \sin(\vartheta) e^{i\phi}|{\Phi_A}{\Phi_B}{\Phi_C}\rangle\right),
\end{equation}

where
\begin{equation}\label{eq:phii1}
\begin{aligned}
    |\Phi_i\rangle &= \cos(\theta_i)|0\rangle + \sin(\theta_i)|1\rangle, \\
    &\quad  \quad i = A, B, C.
\end{aligned}
\end{equation}

and
\begin{equation} \label{eq:p}
P = \left(1 + 2\cos(\vartheta) \sin(\vartheta) \cos(\theta_A) \cos(\theta_B) \cos(\theta_C) \cos(\phi)\right). \\
\end{equation}

\begin{figure}[H]
	\begin{subfigure}{0.42\textwidth}
		\centering
		\includegraphics[width=\textwidth]{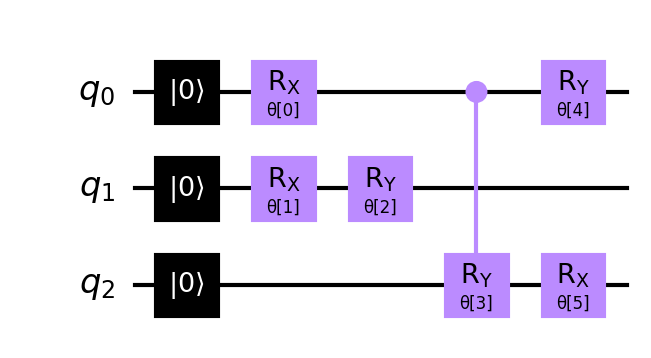}
		\caption{}
		\label{fig:bi part 1 qubit}
	\end{subfigure}
	\begin{subfigure}{0.3\textwidth}
		\centering
		\includegraphics[width=\textwidth]{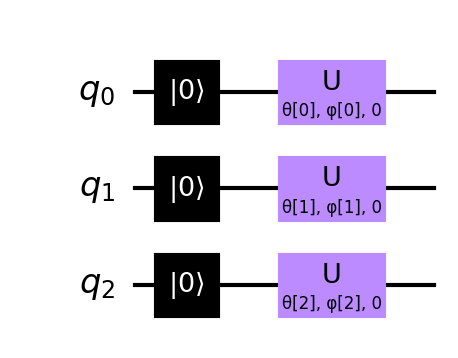}
		\caption{}
		\label{fig:separable 3 qubit } 
	\end{subfigure}
	\caption{Quantum state circuit preparation. (1) An example of a Bi-partite entangled state circuit for three qubits. (2) Separable state circuit.}
	\label{fig:image2}
\end{figure}

The required data for bi-partite state section is assembled from circuit Fig.~\ref{fig:bi part 1 qubit}. We have three possible combinations for this strategy A|BC, B|AC, C|AB (only one of three bi-partite state combinations is explained B|AC). Figure ~\ref{fig:separable 3 qubit } illustrates how we prepare three-qubit separable product states.

\subsection{Entangling Capability and Expressibility }

The entangling capability of a parameterized quantum circuit (PQC) quantifies its ability to generate entanglement between quantum states. This property is crucial for achieving quantum advantage, as entangled states play a central role in quantum algorithms. One method to measure entangling capability is through the Meyer-Wallach entanglement measure~\cite{Meyer11}, which uses average linear entropy. According to Ref.~\cite{sim2019expressibility}, the measure of entanglement of a multi-qubit system is expressed as the average entanglement of each qubit with the rest of the system, as follows:
 
\begin{equation}\label{ent}
	\begin{aligned}
		Q(|\psi \rangle) = 2[1-1/n\sum_{k=1}^{N} tr(\rho_k^2)].
	\end{aligned}
\end{equation}
It can be seen that the value of $Q(\psi)$ depends on the purity of the subsystem states. For instance, a maximally entangled state $Q(\frac{|0,0+|1,1\rangle)}{\sqrt{2}})=1$, while a separable state $Q(|0,0\rangle)=0$.

In this study, we examined 19 (see \ref{fig:all}) distinct VQCs to evaluate and optimize their efficiency. By testing a diverse set of circuit architectures, we identified configurations that maximized the performance in generating entangled states. This comparative approach allowed us to pinpoint the VQC design yielding the highest efficiency for our objectives.

\begin{figure}[H]\label{1234}
	\centering
	\includegraphics[width=0.85\linewidth]{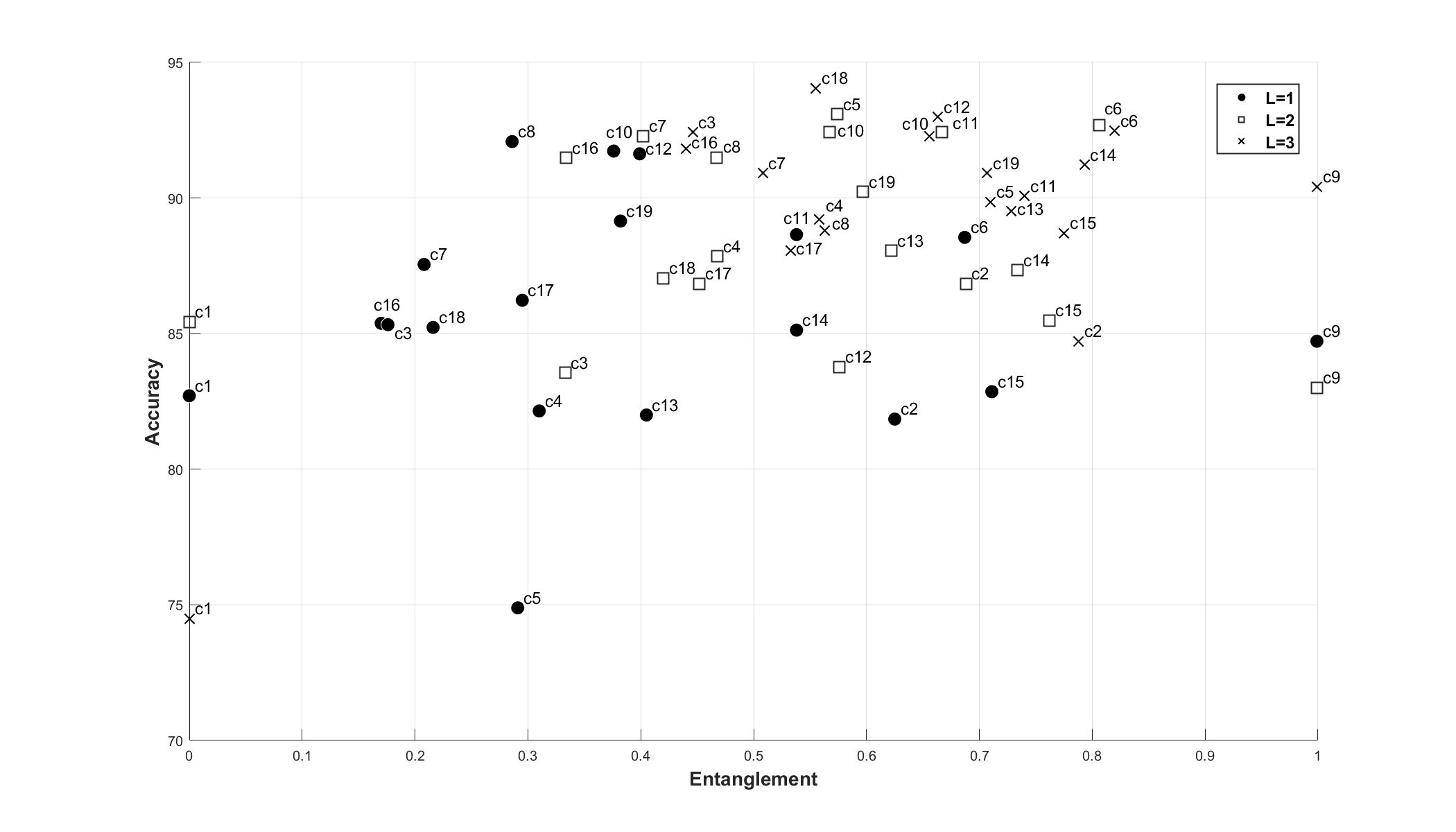}
	\caption{Dispersion of different quantum circuits according to entanglement and accuracy in QSVM.}
	\label{fig:cirplot3}
\end{figure}

According to Figure \ref{fig:cirplot3}, it can be seen that in the one blocks, we need a smaller amount of entanglement for the quantum circuit in order to reach a high accuracy value. In this case, it shows that the increase of blocks will be accompanied by an increase in cost and space for the quantum circuit, which in this figure shows that this increase in the block and volume of the quantum circuit will not necessarily increase the amount of accuracy in quantum processing. The accuracy and entanglement capability values for circuits \emph{C1} through \emph{C19} (see Fig.\ref{fig:all}) are provided in Tab.~\ref{tab:entanglmant} 

\begin{figure}[H]
	\centering
	\includegraphics[width=0.85\linewidth]{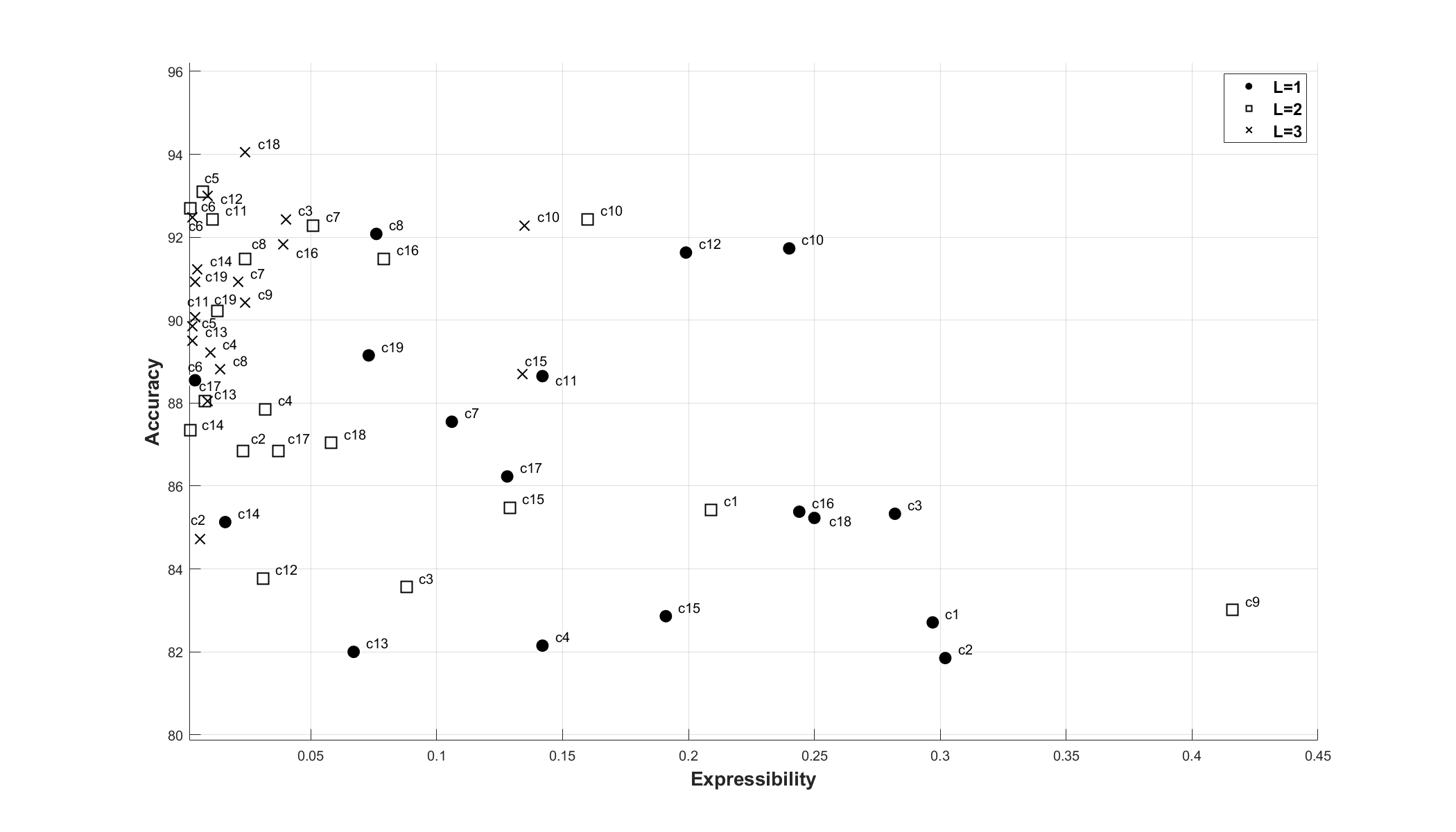}
	\caption{Dispersion of different quantum circuits according to experisibility and accuracy in QSVM.}
	\label{fig:cirplot2}
\end{figure}

PQC designs can generate pure quantum states, and the overlap of these states across the entire Hilbert space is measured by a quantity known as expressibility. To quantify expressibility, we calculate the fidelity of the state distribution obtained from the PQC and compare it with the fidelity distribution of Haar random states:

\begin{equation}\label{exp}
	\begin{aligned}
		EXP= D_{k}(P_{PQC}(F,\theta) || P_{Haar}(F)).
	\end{aligned}
\end{equation}
Here, $D_k$ is the Kullback-Leibler divergence and $P_{PQC}(F,\theta)$ represents the estimated probability distribution of fidelities obtained by sampling states generated from a PQC. 
For an ensemble of Haar-random states, the probability density function of fidelities is given by $ P_{Haar}(F)=(N-1)(1-F)^{(N-2)}$, where F and N are the fidelity and dimension of the Hilbert space, respectively. Expressibility measures how well the state distribution from a PQC matches that of Haar random states, indicating the extent to which the states span the Hilbert space. High expressibility or low Kullback-Leibler divergence values indicate how effectively the PQC can generate all possible quantum states within the Hilbert space (For further details and an example involving a single-qubit state, see Ref.~\cite{sim2019expressibility}).
Figure \ref{fig:cirplot2} illustrates that as the number of blocks in a quantum circuit increases, expressibility also rises. This means the quantum states generated by the circuits cover a broader region of the Hilbert space. Consequently, the accuracy of QSVM in detecting quantum states improves with increased expressibility. Also, the values of accuracy and expressibility for circuits \emph{C1} through \emph{C19} (see Fig.\ref{fig:all}) are provided in Tab.~\ref{tab:Expressibility} 

\begin{table}[H]
	\centering
	\begin{tabular}{|c|c|c|c|c|}
		\hline
		\textbf{Circuit} & \textbf{Accuracy } & \textbf{Entangling Capability} & \textbf{Mean} & \textbf{Variance}\\ \hline
		C1  & 74.49 & 0.000 & 1.30e-32 & 3.84e-68 \\ \hline
		C2  & 81.85 & 0.302 & 0.620 & 6.16e-06  \\ \hline
		C3  & 92.43 & 0.040 & 0.172 & 5.31e-06 \\ \hline
		C4  & 82.15 & 0.142 & 0.311 & 1.11e-05 \\ \hline
		C5  & 74.89 & 0.056 & 0.290 & 1.23e-05\\ \hline
		C6  & 92.48 & 0.820 & 0.684 & 6.52e-06\\ \hline
		C7  & 90.92 & 0.508 & 0.211 & 3.96e-06\\ \hline
		C8  & 88.81 & 0.563 & 0.289 & 1.12e-05\\ \hline
		C9  & 84.72 & 0.788 & 0.381 & 6.14e-06\\ \hline
		C10 & 92.28 & 0.656 & 0.374 & 9.33e-06\\ \hline
		C11 & 90.42 & 0.999 & 0.538 & 3.76e-06\\ \hline
		C12 & 92.99 & 0.663 & 0.389 & 4.12e-06 \\ \hline
		C13 & 89.51 & 0.728 & 0.407 & 5.60e-06\\ \hline
		C14 & 91.22 & 0.793 & 0.548 & 4.31e-06\\ \hline
		C15 & 88.70 & 0.775 & 0.711 & 3.64e-06\\ \hline
		C16 & 91.83 & 0.440 & 0.171 & 6.17e-06\\ \hline
		C17 & 86.23 & 0.128 & 0.293 & 5.17e-06\\ \hline
		C18 & \textbf{94.05} & \textbf{0.555} & 0.219 & 2.97e-06 \\ \hline
		C19 & 90.92 & 0.707 & 0.386 & 8.91e-06\\ \hline
	\end{tabular}
	\caption{We utilized various circuits, ranging from C1 to C19, as detailed in the Appendix C, and conducted a comparison of their Accuracy and Entangling Capability.}
	\label{tab:entanglmant}
\end{table}

\begin{table}[H]
	\centering
	\begin{tabular}{|c|c|c|c|c|c|c|c|}
		\hline
		\textbf{Circuit} & \textbf{Accuracy } & \textbf{Expressibility } &  \textbf{Mean} & \textbf{Variance} \\ \hline
		C1  & 74.49 & 0.211 & 0.294 & 1.37e-04\\ \hline
		C2  & 84.72 & 0.006 & 0.295 & 9.12e-05\\ \hline
		C3  & 92.43 & 0.040 & 0.260 & 2.87e-04 \\ \hline
		C4  & 89.21 & 0.010 & 0.138 & 4.49e-05\\ \hline
		C5  & 89.86 & 0.003 & 0.059 & 2.97e-05\\ \hline
		C6  &92.48 & 0.003 & 0.004 & 5.02e-07\\ \hline
		C7  & 90.92 & 0.021 & 0.115 & 5.37e-05 \\ \hline
		C8  & 88.81 & 0.014 & 0.07 & 4.97e-05\\ \hline
		C9  & 90.42 & 0.024 & 0.670 & 6.40e-04 \\ \hline
		C10 & 92.28 & 0.135 & 0.242 & 1.90e-04\\ \hline
		C11 & 90.07 & 0.004 & 0.145 & 1.13e-04\\ \hline
		C12 & 92.99 & 0.009 & 0.182 & 1.61e-04 \\ \hline
		C13 & 89.51 & 0.003 & 0.063 & 2.15e-05\\ \hline
		C14 & 91.22 & 0.005 & 0.014 & 5.39e-06\\ \hline
		C15 & 88.70 & 0.134 & 0.184 & 9.25e-05\\ \hline
		C16 & 91.83 & 0.039 & 0.255 & 9.58e-05 \\ \hline
		C17 & 88.05 & 0.009 & 0.138 & 6.27e-05\\ \hline
		C18 & \textbf{94.05} & \textbf{0.024} & 0.248 & 1.38e-04\\ \hline
		C19 & 90.92 & 0.004 & 0.071 & 4.60e-05 \\ \hline		
	\end{tabular}
	\caption{We utilized various circuits, ranging from C1 to C19, as detailed in the Appendix C, and conducted a comparison of their Accuracy and  Expressibility.}
	\label{tab:Expressibility}
\end{table}

\section{Investigating IBMQ through Experimental Trials}

The IBM quantum computer is widely recognized as a leader in the quantum computing domain, contributing significantly to advancements in quantum hardware and software. By leveraging quantum phenomena such as superposition, entanglement, and interference, it enables the efficient and accurate resolution of complex problems beyond the capabilities of classical computers.
This innovation has driven exploration in quantum algorithms and has led to applications across diverse fields, including chemistry, finance, logistics, and artificial intelligence.
With the introduction of Q System One, IBMQ has made strides toward commercializing quantum computing, unlocking opportunities to tackle previously unsolvable challenges. As industries increasingly recognize the transformative potential of quantum computing, IBMQ's pioneering role is poised to shape the future of computation and machine learning~\cite{ibm}.

IBM currently offers cloud-based quantum computing services to developers, researchers, and businesses through its Quantum Experience platform. In this study, we report the results of experiments conducted using both IBM's noisy simulators and real quantum computing systems. First, the Qiskit runtime "ibm-quantum" channel was selected, and the available backends were identified. Among them, "ibm-perth," "ibm-lagos," and "ibm-nairobi" are seven-qubit quantum processors, while "simulator-stabilizer," "simulator-mps," "simulator-extended-stabilizer," "ibmq-qasm-simulator," and "simulator-statevector" are accessible simulator services, each with distinct error maps in computational processes.

\begin{figure} [H]
	\centering
	\begin{subfigure}[H]{0.3\textwidth}
		\centering
		\includegraphics[width=\textwidth]{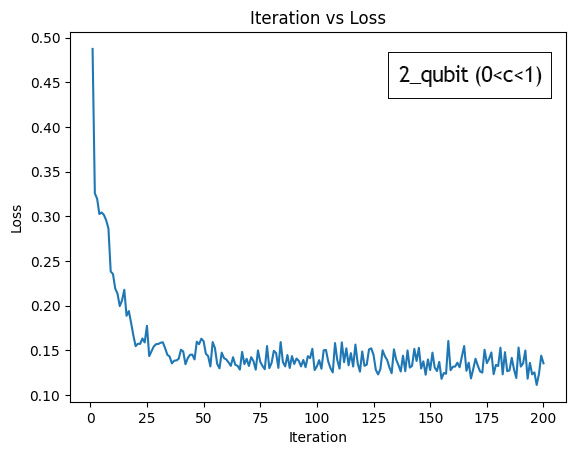}
		\caption{}
		\label{fig:loss (1)}
	\end{subfigure}
	\hfill
	\begin{subfigure}[H]{0.3\textwidth}
		\centering
		\includegraphics[width=\textwidth]{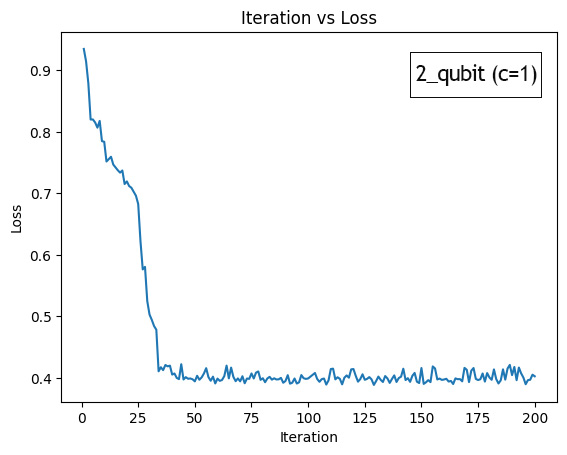}
		\caption{}
		\label{fig:loss (2)}
	\end{subfigure}
	\hfill
	\begin{subfigure}[H]{0.3\textwidth}
		\centering
		\includegraphics[width=\textwidth]{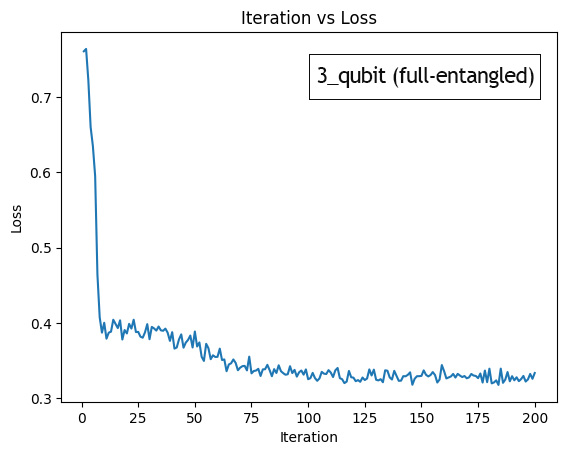}
		\caption{}
		\label{fig:loss (3)}
	\end{subfigure}
	\hfill
	\begin{subfigure}[H]{0.3\textwidth}
		\centering
		\includegraphics[width=\textwidth]{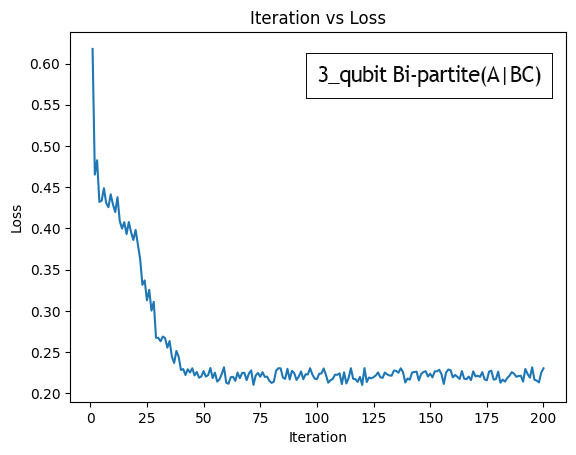}
		\caption{}
		\label{fig:loss (4)}
	\end{subfigure}
	\hfill
	\begin{subfigure}[H]{0.3\textwidth}
		\centering
		\includegraphics[width=\textwidth]{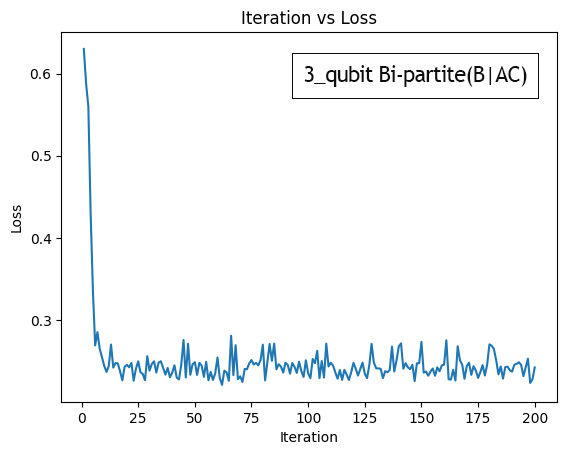}
		\caption{}
		\label{fig:loss (5)}
	\end{subfigure}
	\hfill
	\begin{subfigure}[H]{0.3\textwidth}
		\centering
		\includegraphics[width=\textwidth]{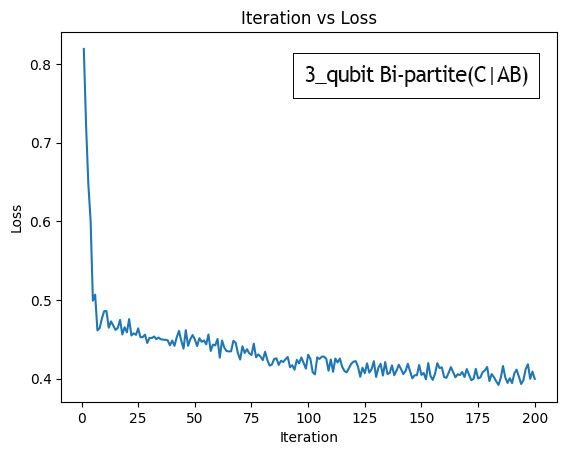}
		\caption{}
		\end{subfigure}
	\caption{ The optimization levels were computed using the SPSA optimizer with 200 iterations, ordering from the first data dimension as shown in Tab.~\ref{tab:Test accuracy}.}	\label{loss1221}
\end{figure}

Our method necessitated classical optimization, so we incorporated session definitions in our code. The number of circuit shots was set to 8192. During training data sampling, the data dimension was reduced to $n = 2$  with a size of  $N=4$ to mitigate decoherence effects.Twenty test datasets were generated with random $\theta $ values using the circuit shown in Fig.~\ref{fig:2 qubit}. The Simultaneous Perturbation Stochastic Approximation (SPSA) minimization tool was chosen as it aligns well with the random nature of quantum states and the data sampling ~\cite{spsa,spsa1}. The optimization steps up to the minimum stage with 200 iterations for Table  ~\ref{tab:Test accuracy} are shown in Figure ~\ref{loss1221}. The elapsed time for each circuit's runtime process is presented in table~\ref{tab:time-usage}.

\begin{table}[ht]
\centering
\begin{tabular}{||c c c c c||}
 \hline
 circuit & C1 & C2 & C3 & C4 \\ [0.5ex]
 \hline\hline
 $t_s$ & 11 & 3 & 3 & 3 \\ [1ex]
 \hline
\end{tabular}
\caption{Real quantum processor time usage at one stage of the performance.}
\label{tab:time-usage}
\end{table}

Experiments were run on the \texttt{IBM-Perth} seven-qubit processor, as well as the \texttt{ibmq-qasm-simulator}. This work incorporates the latest updates from \texttt{IBMQ} services and provider information~\cite{ibm_quantum_docs}.

\begin{table}[ht]
   \centering
   \scriptsize
   \begin{tabular}{|c|c|c|c|c|c|c|}
      \hline
      \multirow{2}{*}{\rotatebox[origin=c]{90}{\parbox{3cm}{Data dimension}}}
      & \multirow{1}{*}{\rotatebox[origin=c]{90}{\parbox{3cm}{IBMQ-\hspace{1pt}Real\\ \hspace{1pt}System}}}
      & \multirow{2}{*}{\rotatebox[origin=c]{90}{\parbox{3cm}{IBMQ-\hspace{1pt}Noisy\\Simulator}}}
      & \multicolumn{4}{c|}{Qiskit Aer-Simulator} \\
      \cline{4-7}
      &&& \rotatebox[origin=c]{90}{\parbox{3cm}{\centering VQC(\#1)}} & \rotatebox[origin=c]{90}{\parbox{3cm}{\centering VQC(\#2)}} & \rotatebox[origin=c]{90}{\parbox{3cm}{\centering VQC(\#3)}} & \rotatebox[origin=c]{90}{\parbox{3cm}{\centering VQC(\#4)}} \\
      \hline
      2-qubit ,(0$<$c$<$1) &  90.01\% & 92.1\% & 97.4\% & 92.5\% & 96.6\% & 95.3\% \\[1ex]
      \hline
      2-qubit ,(c$=$1) &  93.2\% & 92.04\%  & 100\% & 94.1\% & 98.1\% & 97.6\%  \\[1ex]
      \hline
      3-qubit,(full-entangled)  & - & 89.6\% & 96.2\% & 91.8\% & 93.8\% & 92.4\% \\[1ex]
      \hline
      3-qubit,Bi-partite(A|BC)  & - & 85.2\% & 92.7\% & 86.3\% & 91.0\% & 89.9\%  \\[1ex]
      \hline
      3-qubit,Bi-partite(B|AC)  & - & 83.1\% & 91.5\% & 85.7\% & 89.4\% & 88.6\%  \\[1ex]
      \hline
      3-qubit,Bi-partite(C|AB)  & - & 87.7\% & 94.1\% & 88.4\% & 91.7\% & 93.1\%  \\[1ex]
      \hline
   \end{tabular}
   \caption{Test accuracy for presented approach.}
   \label{tab:Test accuracy}
\end{table}

\begin{table}
	\centering
	\begin{tabular}{|c| c| c |c |c|}
		\hline
		qubit & data size & test data &  mean error & variance  \\ [0.5ex]
		\hline\hline
		2-qubit & 4000 & 3992 & 0.026 & 1.94e-05 \\ [1ex]
		\hline
		3-qubit & 4000 & 3884 & 0.192 & 4.34e-06 \\ [1ex]
		\hline
		4-qubit & 4000 & 3884 & 0.014 & 4.14e-05 \\ [1ex]
		\hline
	\end{tabular}
	\caption{Data usage for two, three and four qubits. }
	\label{tab:data-usage}
\end{table}

\section{Conclusion}

In conclusion, we successfully distinguished separable quantum states from entangled states using the QSVM method. We meticulously prepared the training and testing datasets from the generated quantum states. For the training dataset, we randomly selected a subset of quantum states, labeling them as either separable or entangled. Similarly, we randomly chose a subset for the testing dataset. Our findings indicate that the QSVM model can accurately predict the labels of these quantum states.
The recommendation model was effectively developed to accurately predict both two-qubit and three-qubit states. Specifically, we utilized a two-qubit quantum circuit across at least four training datasets, resulting in optimal performance for both two-qubit and three-qubit datasets. Furthermore, by increasing the training datasets to eight and employing a three-qubit VQC-SVM model, we observed a significant enhancement in accuracy, particularly for the three-qubit datasets.
To further assess model performance, we tested several VQC circuits to compare their results and identify the model that provided the highest accuracy, as detailed in Tab.~\ref{tab:Test accuracy}. Overall, this study underscores the effectiveness of the QSVM method in classifying quantum states and highlights the importance of circuit choice in optimizing predictive capabilities.

\section{Appendix}
\appendix
\section{Support vector machine}

The fundamental idea behind SVM is to find an optimal hyperplane that maximally separates the data points of different classes in a high-dimensional feature space. SVM is particularly effective for binary classification tasks, where the goal is to separate data points into two distinct classes.
For classification, if the training dataset is linearly separable SVM determines a maximum margin separating hyperplane with its normal vector $w$ the $w^T.\vec{x_i}+b=y$ , which divides two classes with labels $y_i \in\left\{-1,1\right\}$. The two parallel hyperplanes are located at a maximum distance of $2/|\vec{w}|$ apart, with no data points in between them, forming the "hard" margin Eq.~\eqref{eq:hard}. To increase the margin's flexibility, it's possible to allowing some of the training data to encroach on the "soft" margin Eq.~\eqref{eq:soft}.However, to minimize the aforementioned violation, it is advisable to prevent $\xi_i$ from becoming excessively large by imposing a penalty on them within the objective function.

For instance, we can incorporate a term $\sum_{i=1}^{N} \xi_i$ into the objective function. In soft margin The objective function has two terms: one that minimizes $\|w\|^2$ (which maximizes the margin), and another that minimizes $\sum_{i=1}^{N} \xi_i$ which is a measure of how much the constraints for $i = 1,\ldots, N$ , are violated. The parameter $C$ controls the trade-off between the two terms. A high value of $C$ results in the optimizer accepting very few errors, while ($C \rightarrow \infty$), it corresponds to the case where the data is perfectly separable.
\begin{equation}\label{eq:hard}
\begin{aligned}
\min_{w, b}\: & \frac{1}{2\|w\|^2} \\
\text{s.t.} & \quad y_i(w^T \cdot \vec{x_i} + b) \geq 1, \quad \forall i
\end{aligned}
\end{equation}

\begin{equation}\label{eq:soft}
\begin{aligned}
\min_{w, b, c}\: & \frac{1}{2\|w\|^2} + C \sum_{i=1}^{N} \xi_i^2 \\
\text{s.t.} & \quad y_i(w^T \cdot \vec{x_i} + b) \geq (1 - \xi_i), \quad \forall i \\
& \quad \xi_i \geq 0, \quad i = 1, \ldots, N
\end{aligned}
\end{equation}
 Re-writing Eq.~\eqref{eq:hard} and~\eqref{eq:soft} to obtain the Lagrangian in its primal form yields

\begin{equation}\label{eq:dual-primal}
\mathcal{L}(w, b, \alpha) = \frac{||w||^2}{2} - \sum_{i}^N \mu_i y_i (w^T\cdot x_i + b) + \sum_{i}^N \mu_i\ ,
\end{equation}
Finding the hyperplane is equal to solving the Karush–Kuhn–Tucker (KKT) requirements since obtaining the SVM hyperplane is a convex optimization problem (Fletcher 2013). For ~\eqref{eq:dual-primal} the KKT conditions are:
\begin{equation}\label{eq:KKT}
\begin{aligned}
\left\{\begin{array}{l}
\frac{\partial}{\partial \boldsymbol{w}} \mathcal{L}(w, b, \mu)=w-\sum_{i=1}^N \mu_i y_i \boldsymbol{x}_i=0 , \\
\\
\frac{\partial}{\partial b} \mathcal{L}(w, b, \mu)=-\sum_{i=1}^N \mu_i y_i=0 ,
\end{array}\right.
\end{aligned}
\end{equation}
By substituting the expression for$w$ from Eq.~\eqref{eq:dual-primal} into Eq.~\eqref{eq:KKT}, we can enrich and refine the formulation.
This approach facilitates expressing the restrictions in~\eqref{eq:hard} as restraints on the Lagrange multipliers. The data in both the training and test subsets will manifest as the product of the respective vectors involved.
\begin{equation}\label{eq:dualform}
\begin{aligned}
\mathcal{L}(\mu) = \sum_{i=1}^{N} \mu_i - \frac{1}{2} \sum_{i=1}^{N} \sum_{j=1}^{N} \mu_i \mu_j y_i y_j x_i^T x_j , \\
\end{aligned}
\end{equation}
\begin{equation}\label{eq:con1}
\begin{aligned}
y_i\left(\boldsymbol{w}^T \boldsymbol{x}_i+b\right)-1 & \geq 0 \quad &&(i=1,2, \ldots, N) ,
\end{aligned}
\end{equation}
\begin{equation}\label{eq:con2}
\begin{aligned}
\alpha_i & \geq 0 \quad &&(\forall_i) ,
\end{aligned}
\end{equation}
\begin{equation}\label{eq:con3}
\begin{aligned}
\mu_i\left(y_i\left(\boldsymbol{w}^T \boldsymbol{x}_i+b\right)-1\right) & =0 \quad &&(\forall_i) ,
\end{aligned}
\end{equation}

where $\mathcal{L}(\mu)$ stands for the Lagrangian's dual form~\eqref{eq:dualform}. Given the limitations listed, the dual problem can be resolved by minimizing $\mathcal{L}(\alpha)$ with respect to $\alpha$, subject to the constraints given in Eqs.\eqref{eq:con1}-\eqref{eq:con3}.
In conclusion, the decision function turns into:
\begin{equation}\label{eq:decision}
\begin{aligned}
{f}(\boldsymbol{a}) = \text{sgn}\left(\sum_{i=1}^{N} \mu_i y_i \boldsymbol{x}_i^T \boldsymbol{a} + b\right) ,
\end{aligned}
\end{equation}
Note that $\boldsymbol{a}$ includes test data.
The introduction of soft margin SVM is made possible by taking this trade-off into account.Therefore the corresponding KKT complementarity conditions for Eq.\eqref{eq:soft} can be readily converted into Lagrangian in its primal Eq.\eqref{eq:soft dual} and dual Eq.\eqref{eq:soft dual la} form, much like in the separable case:
\begin{equation}\label{eq:soft dual}
\mathcal{L}(\boldsymbol{w}, b, \mu,\xi,\kappa)=\frac{\|\boldsymbol{w}\|^2}{2}+C \sum_{i=1}^N \xi_i^2-\sum_{i=1}^N \mu_i\left[y_i\left(\boldsymbol{w}^T \boldsymbol{x}_i+b\right)-1+\xi_i\right]-\sum_{i=1}^N \kappa_i \xi_i ,
\end{equation}
\begin{equation} \label{eq:soft dual laa}
\mathcal{L}(\mu, C,\lambda) = \frac {1}{C} \sum_{i=1}^{N} \mu_i^2 + \frac{1}{2}\sum_{i,j=1}^{N} y_i y_j \mu_i  \mu_j (x_i^T x_j) - \frac{1}{2\lambda} ,
\end{equation}
The decision function performs after\eqref{eq:soft dual laa} is optimized to get minimum $\alpha$ values Eq.\eqref{eq:decision soft}.
\begin{equation}\label{eq:decision softt}
\begin{aligned}
{f}(\boldsymbol{a}) = \text{sgn}\left(\sum_{i=1}^{N} \mu_i y_i \boldsymbol{x}_i^T \boldsymbol{a} + \frac{1}{\lambda}\right),
\end{aligned}
\end{equation}
where \text{sgn}(·) is a sign function defined by:
\[
\text{{sgn}}(a) =
\begin{cases}
    -1, & \text{{if }} a < 0 ; \\
    1, & \text{{if }} a \geq  0 .
\end{cases}
\]

\section{An extensive look into IBMQ's experiment}

\subsection{Quantum simulator}
The quantum simulations were run on simulator, which is available on
Qiskit and IBM Quantum Services. Each circuit was run with 8192 shots. 

\subsection{Quantum hardware}
The following four 27-qubit super conducting quantum computers available on IBM
Quantum Services were used to run the quantum algorithms: ibmq kolkota, ibmq montreal, ibmq mumbai,
and ibmq auckland. 

\subsection{Software}
The algorithms were implanted in Python 3 $(http://www.python.org)$. The open-source SDK Qiskit $(https://qiskit.org)$ was used to work with quantum algorithms. 
At the time this article was written, IBMQ offered three real seven-qubit systems: "ibm-perth," "ibm-lagos," and "ibm-nairobi".

\subsection{Data availability}
The numerical data generated in this work are available from the corresponding author on reasonable request.
\section{Variational quantum circuits}

\captionsetup[subfigure]{labelformat=empty}
\begin{figure}[H]  
    \centering  
    
    \subfloat[$C1$]{%
        \includegraphics[width=0.16\textwidth]{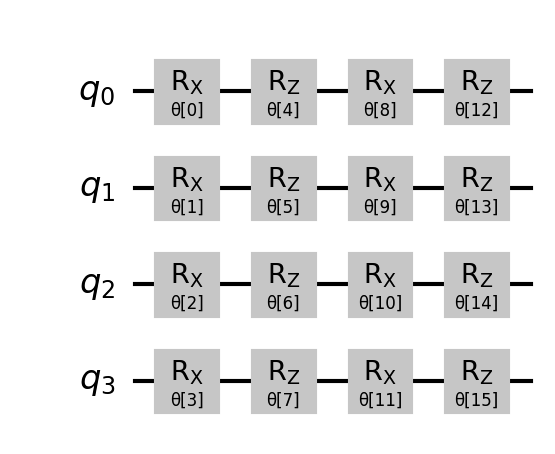}%
        \label{fig:1}%
    }\hfill  
    \subfloat[$C2$]{%
        \includegraphics[width=0.32\textwidth]{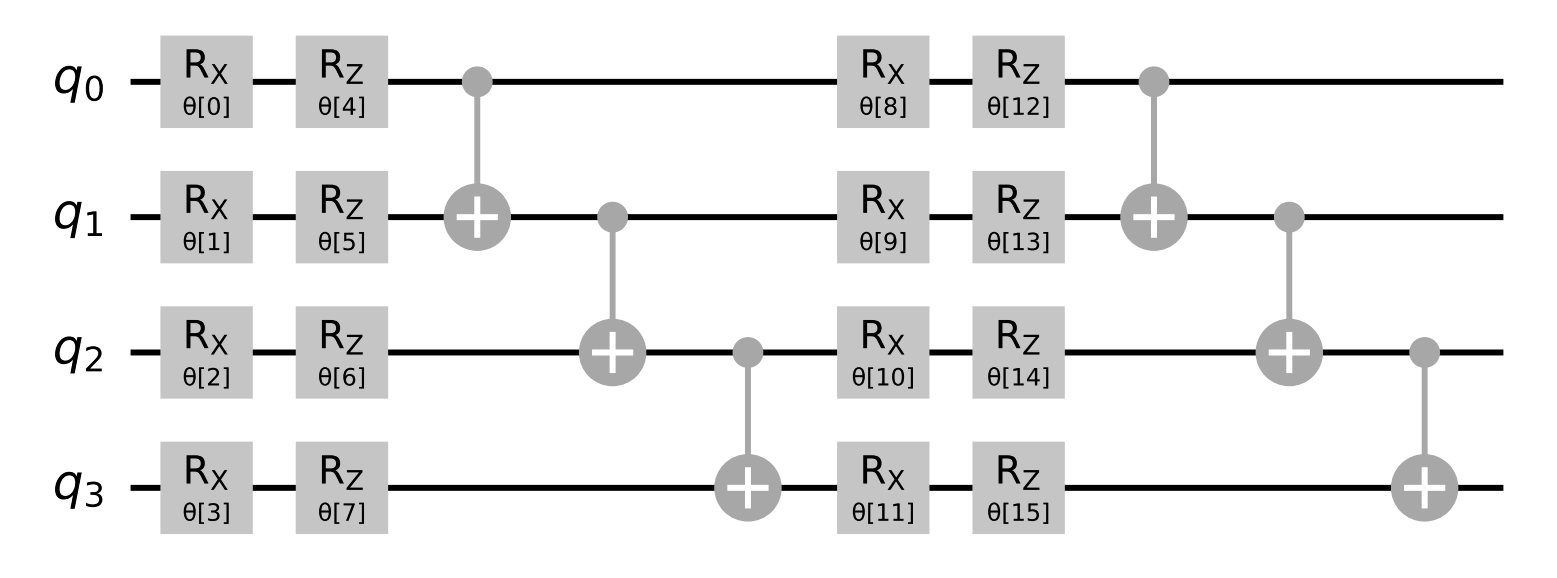}%
        \label{fig:2}%
    }\hfill  
    \subfloat[$C3$]{%
        \includegraphics[width=0.32\textwidth]{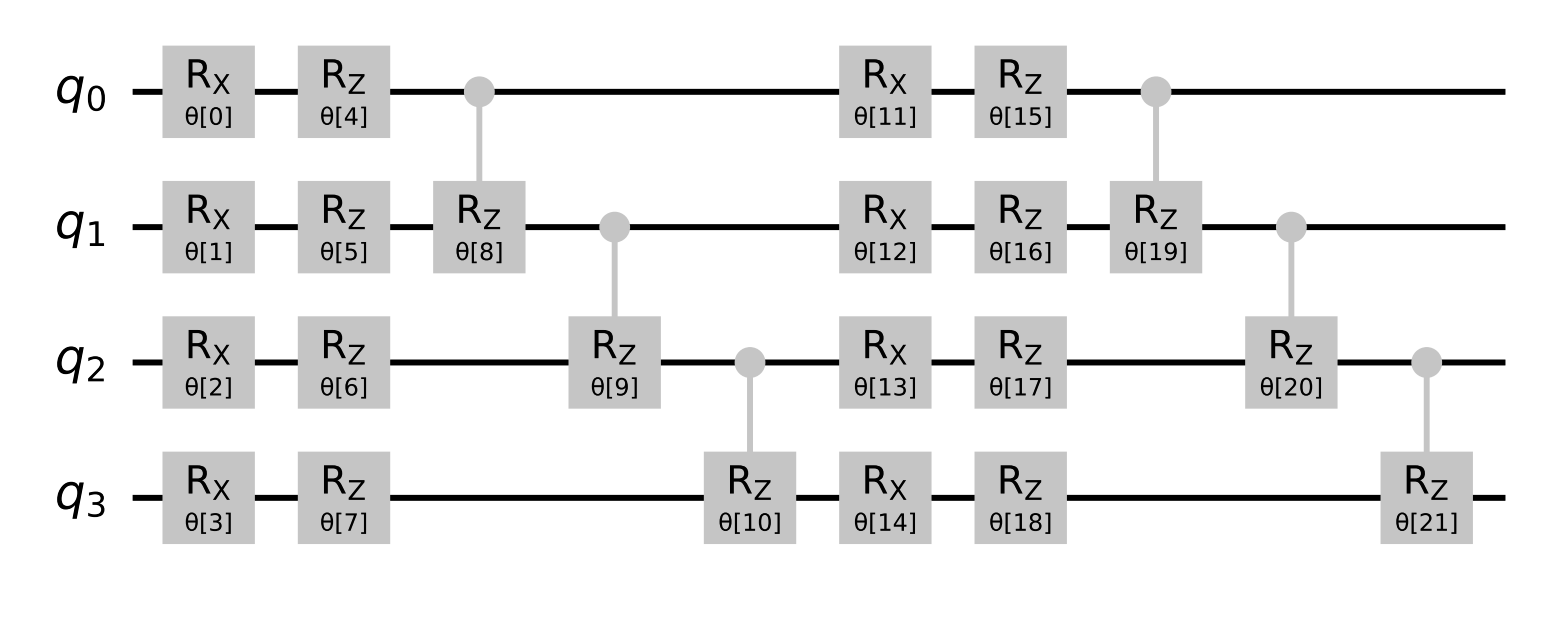}%
        \label{fig:3}%
    }\\  
    
    \subfloat[$C4$]{%
        \includegraphics[width=0.32\textwidth]{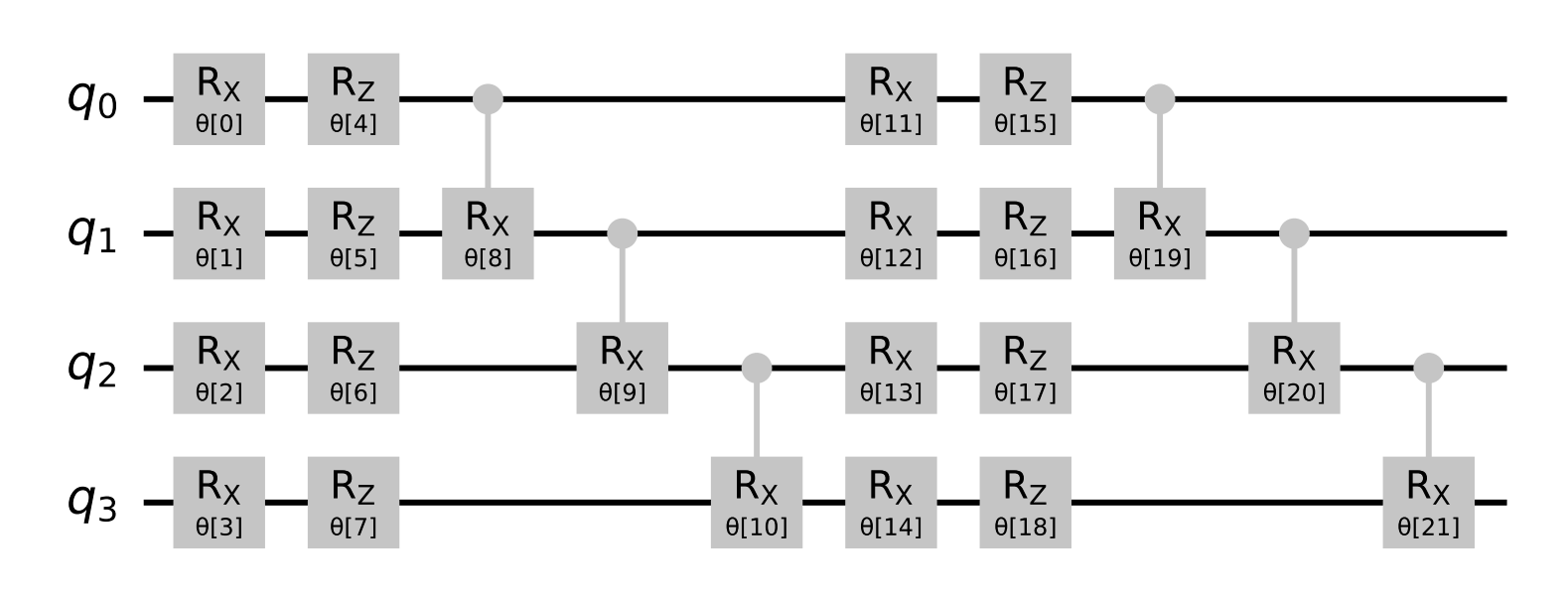}%
        \label{fig:4}%
    }\hfill  
    \subfloat[$C5$]{%
        \includegraphics[width=0.66\textwidth]{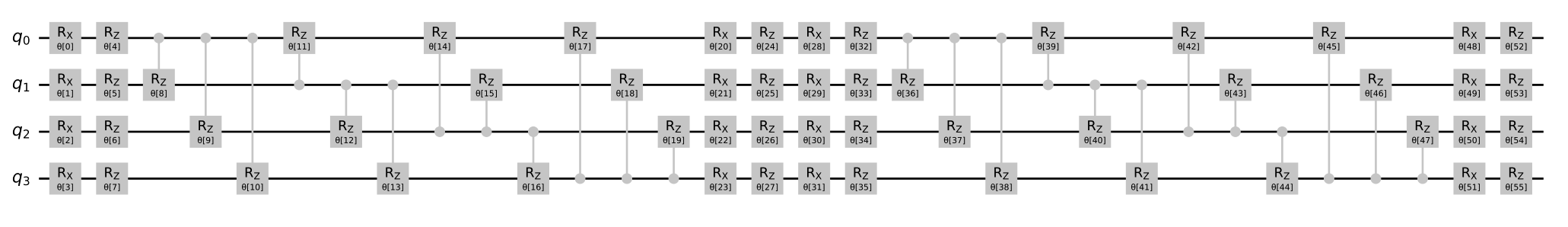}%
        \label{fig:5}%
    }\\

    \subfloat[$C6$]{%
        \includegraphics[width=0.66\textwidth]{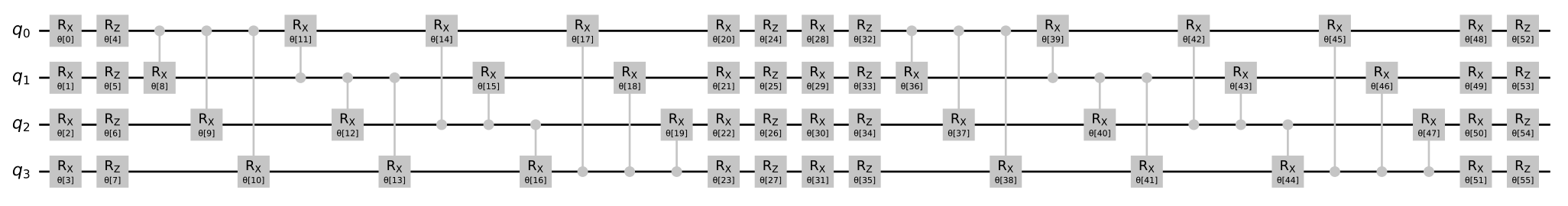}%
        \label{fig:6}%
    }\hfill  
    \subfloat[$C7$]{%
        \includegraphics[width=0.32\textwidth]{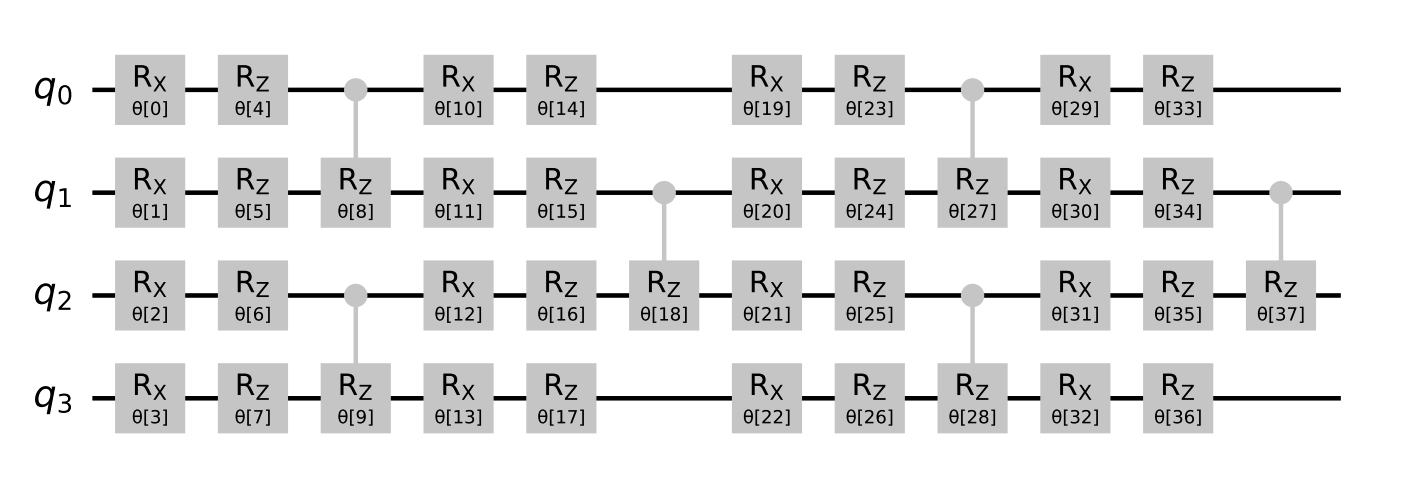}%
        \label{fig:7}%
    }\\
 
    \subfloat[$C8$]{%
        \includegraphics[width=0.32\textwidth]{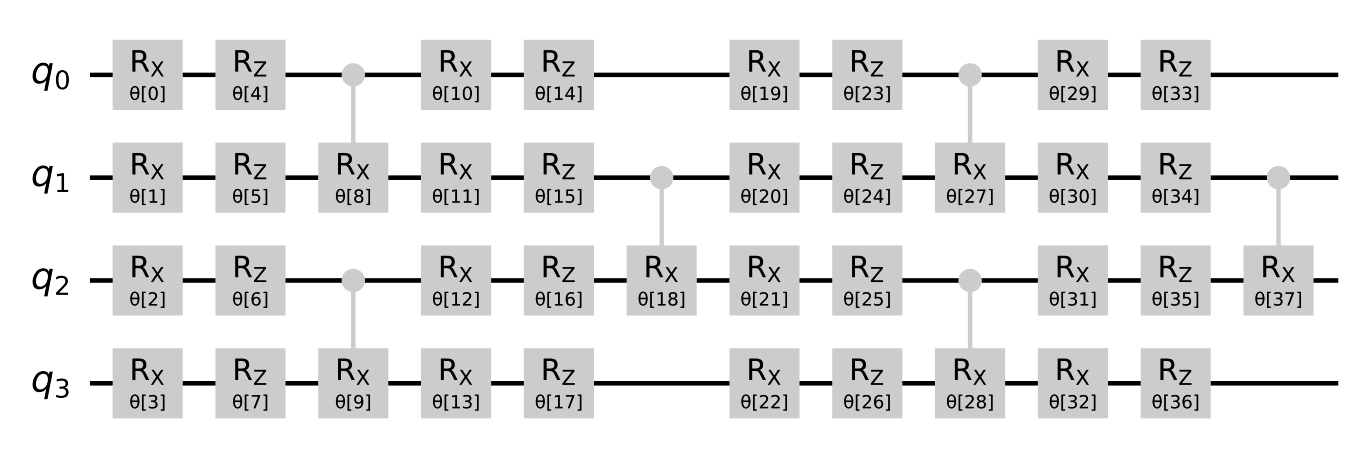}%
        \label{fig:8}%
    }\hfill  
    \subfloat[$C9$]{%
        \includegraphics[width=0.32\textwidth]{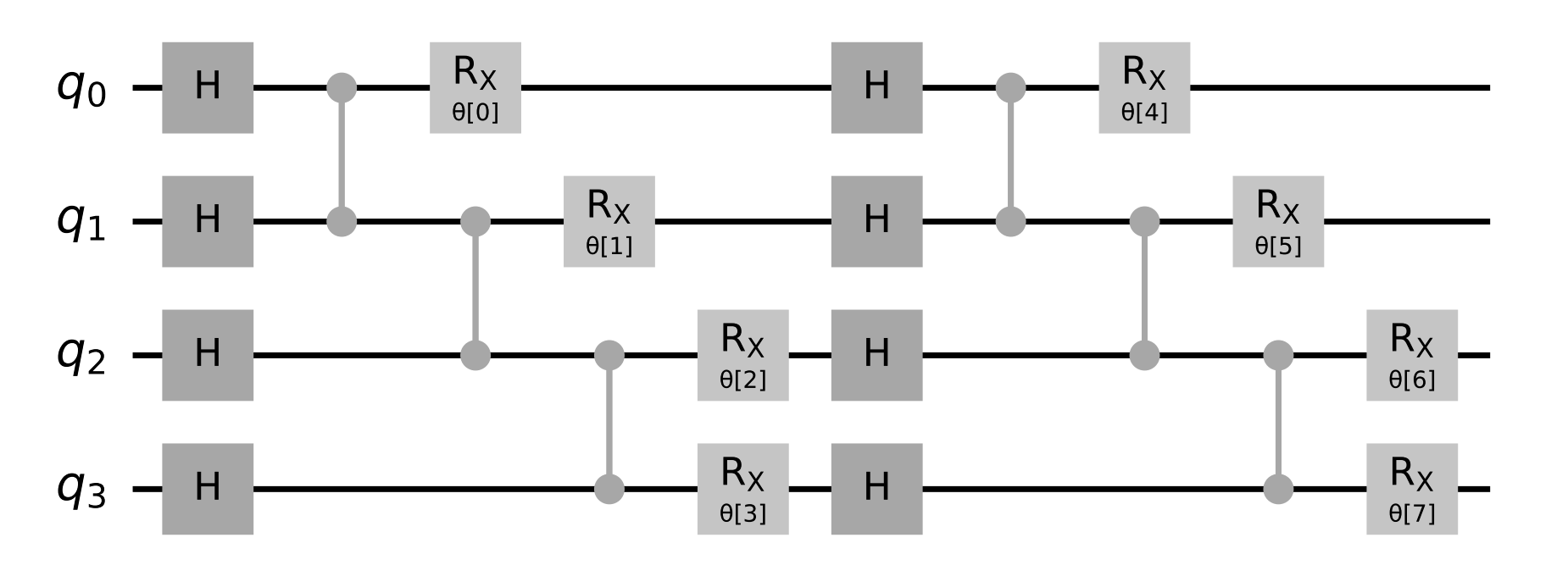}%
        \label{fig:9}%
    }\hfill  
    \subfloat[$C10$]{%
        \includegraphics[width=0.32\textwidth]{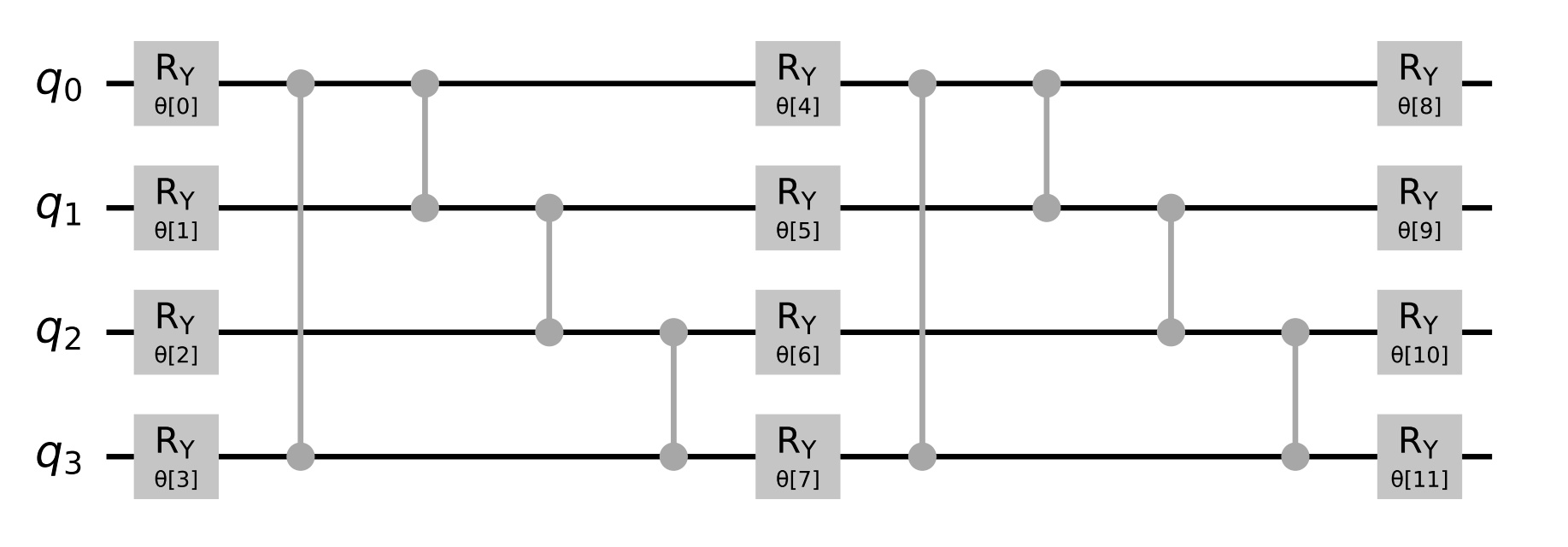}%
        \label{fig:10}%
    }\\  

    \subfloat[$C11$]{%
        \includegraphics[width=0.32\textwidth]{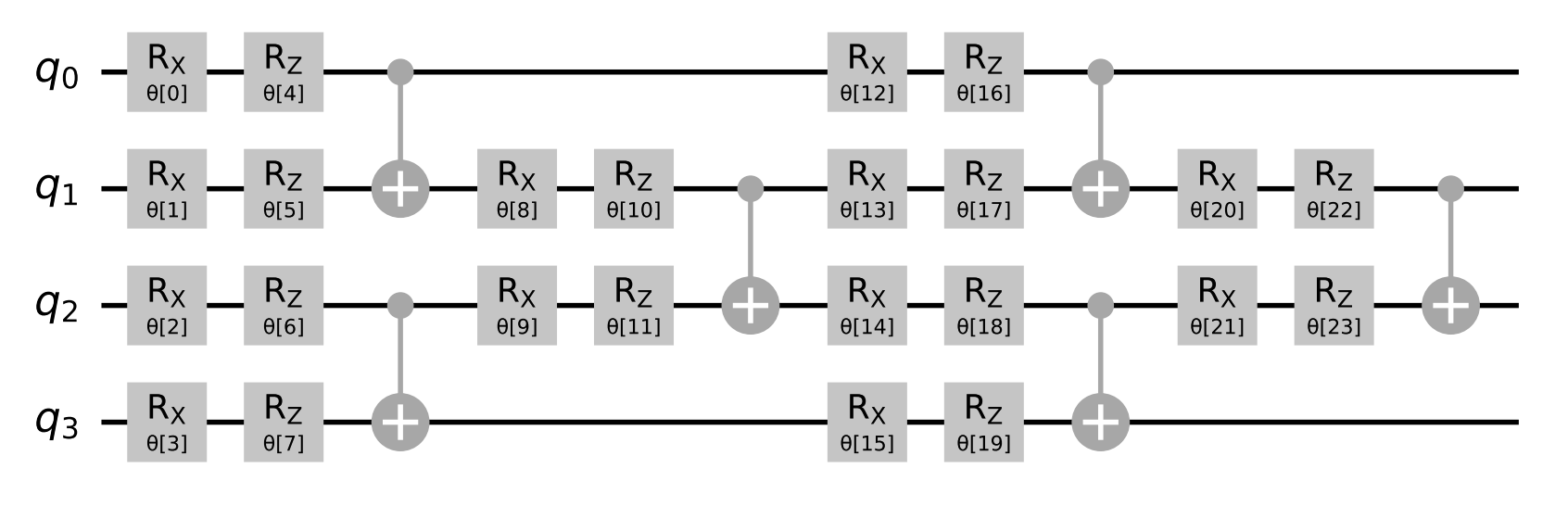}%
        \label{fig:11}%
    }\hfill  
    \subfloat[$C12$]{%
        \includegraphics[width=0.32\textwidth]{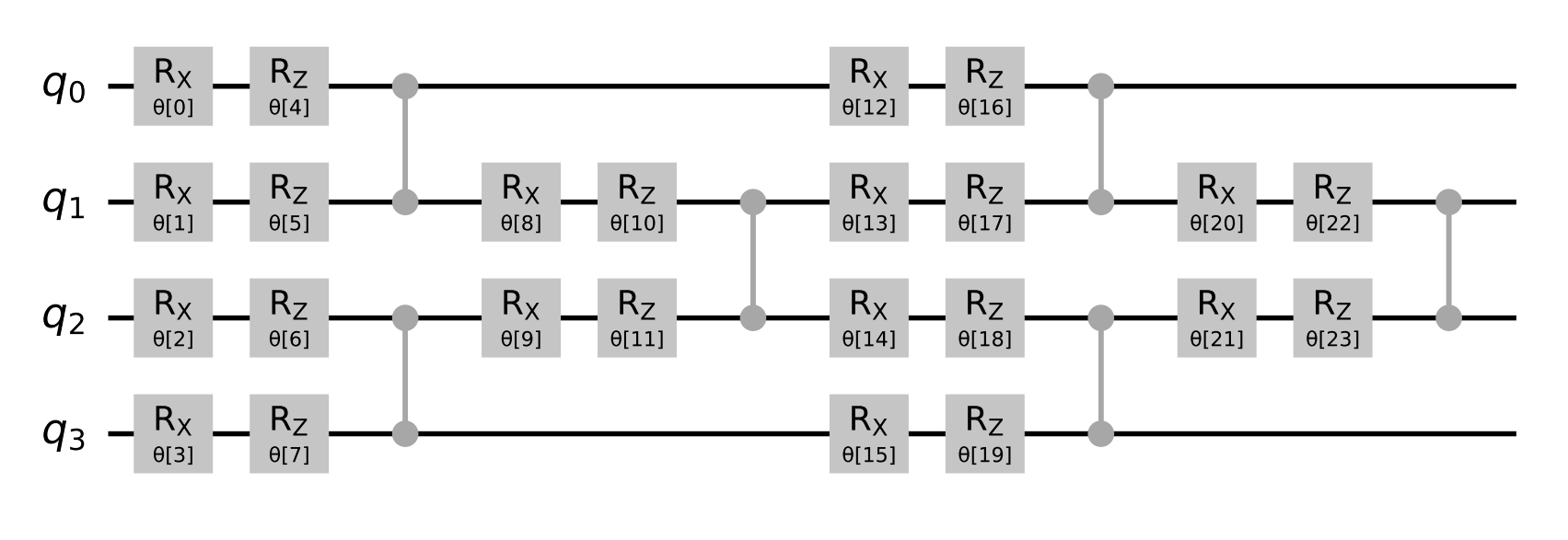}%
        \label{fig:12}%
    }\hfill  
    \subfloat[$C13$]{%
        \includegraphics[width=0.32\textwidth]{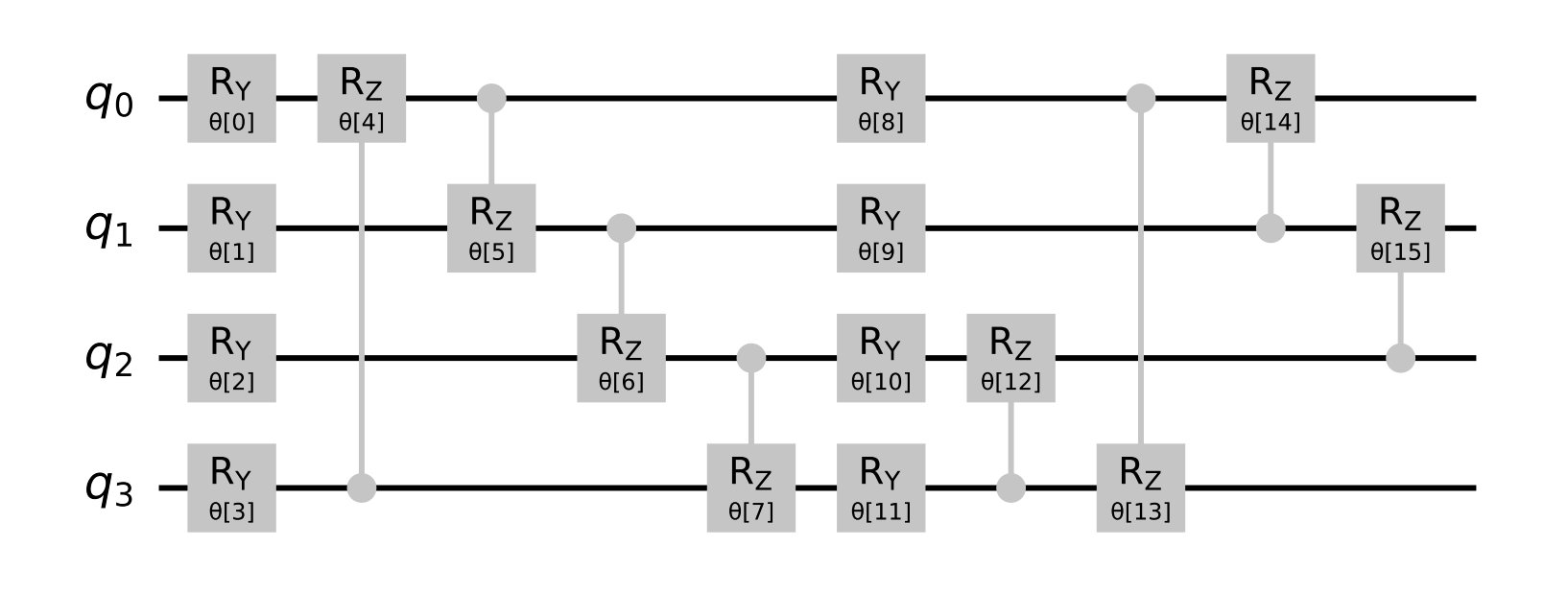}%
        \label{fig:13}%
    }\\ 

    \subfloat[$C14$]{%
        \includegraphics[width=0.32\textwidth]{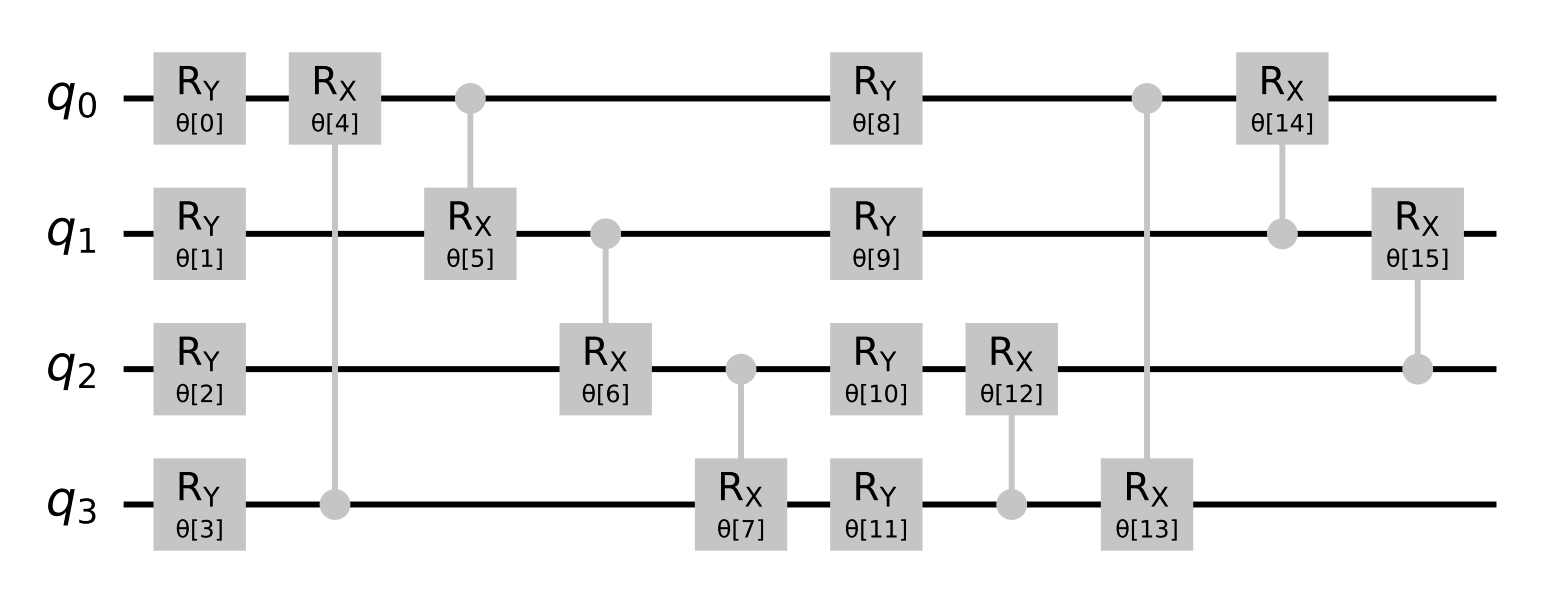}%
        \label{fig:14}%
    }\hfill  
    \subfloat[$C15$]{%
        \includegraphics[width=0.32\textwidth]{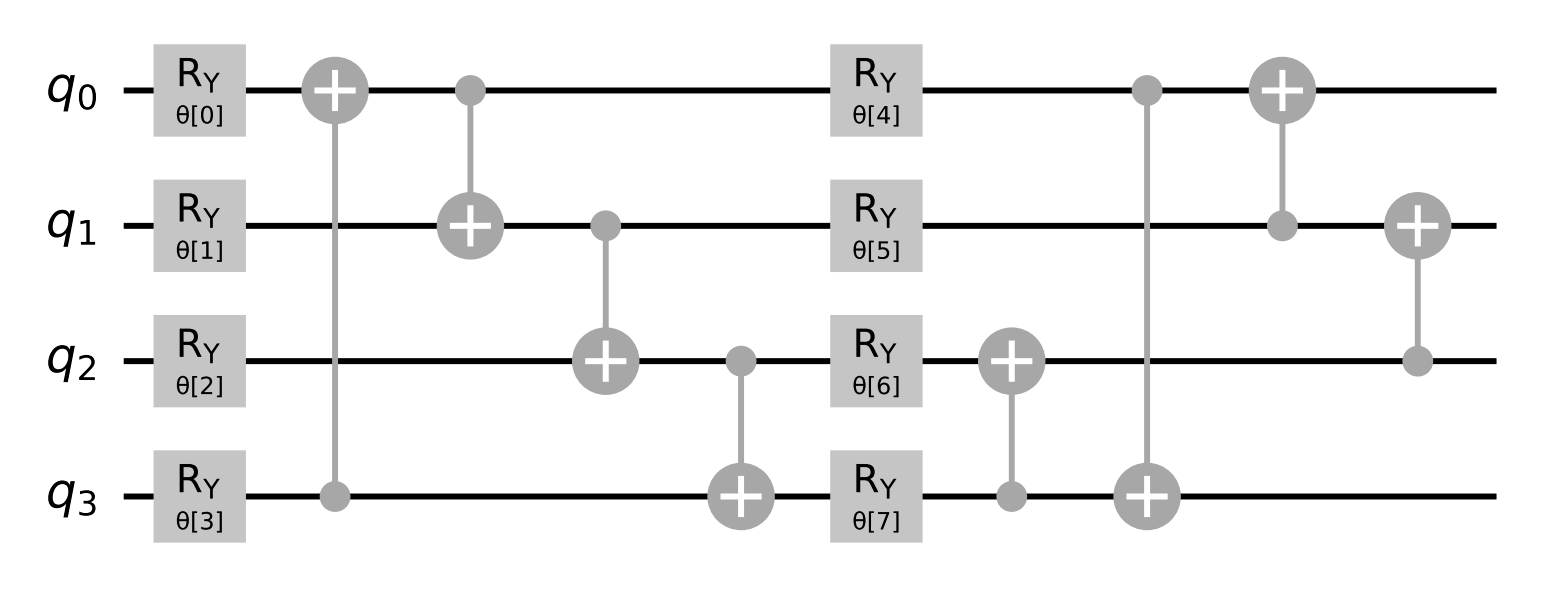}%
        \label{fig:15}%
    }\hfill  
    \subfloat[$C16$]{%
        \includegraphics[width=0.32\textwidth]{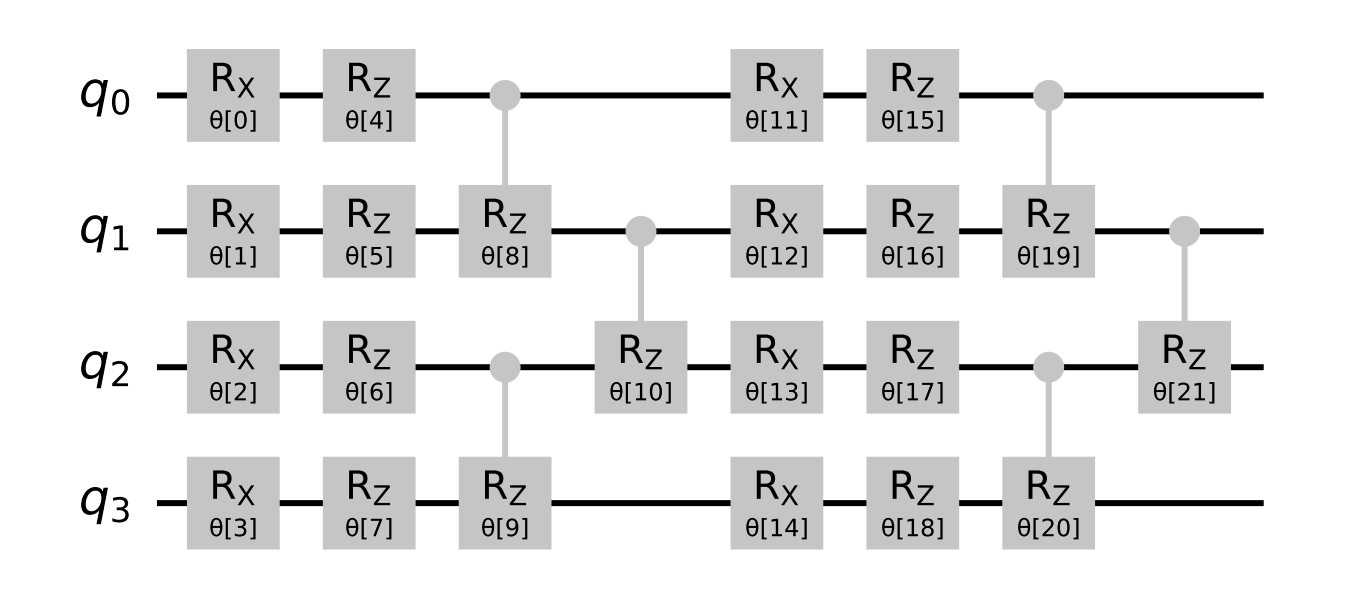}%
        \label{fig:16}%
    }\\

    \subfloat[$C17$]{%
        \includegraphics[width=0.25\textwidth]{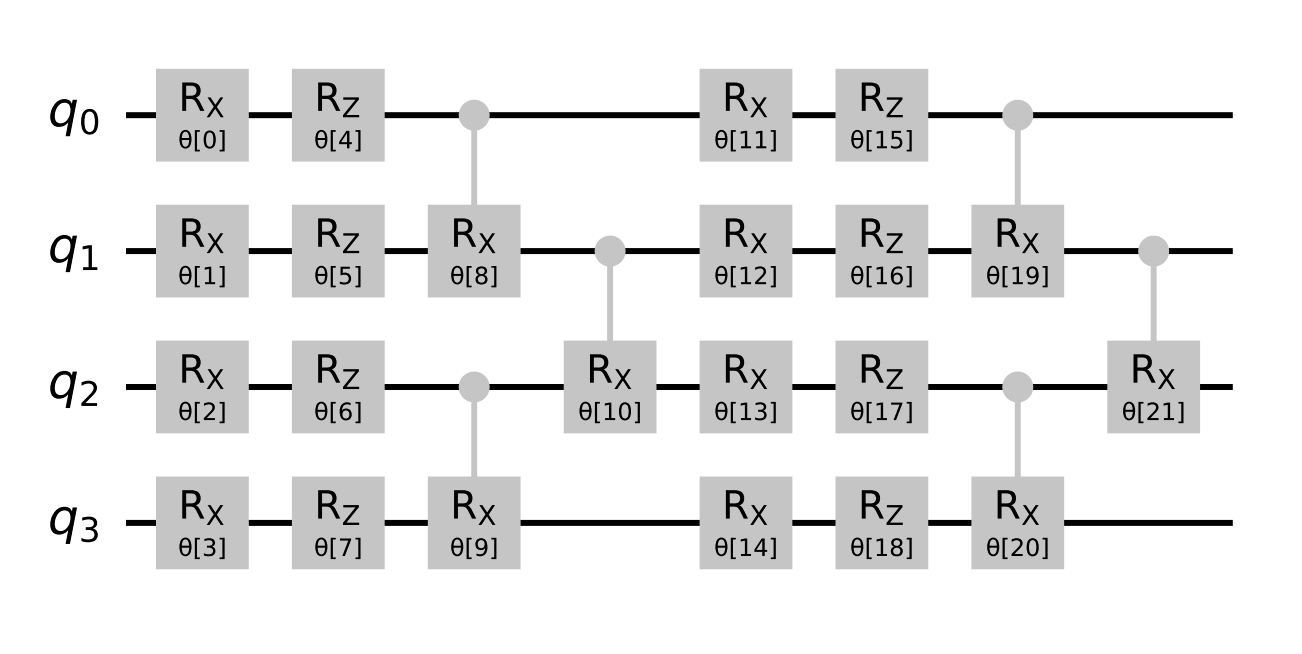}%
        \label{fig:17}%
    }\hfill  
    \subfloat[$C18$]{%
        \includegraphics[width=0.32\textwidth]{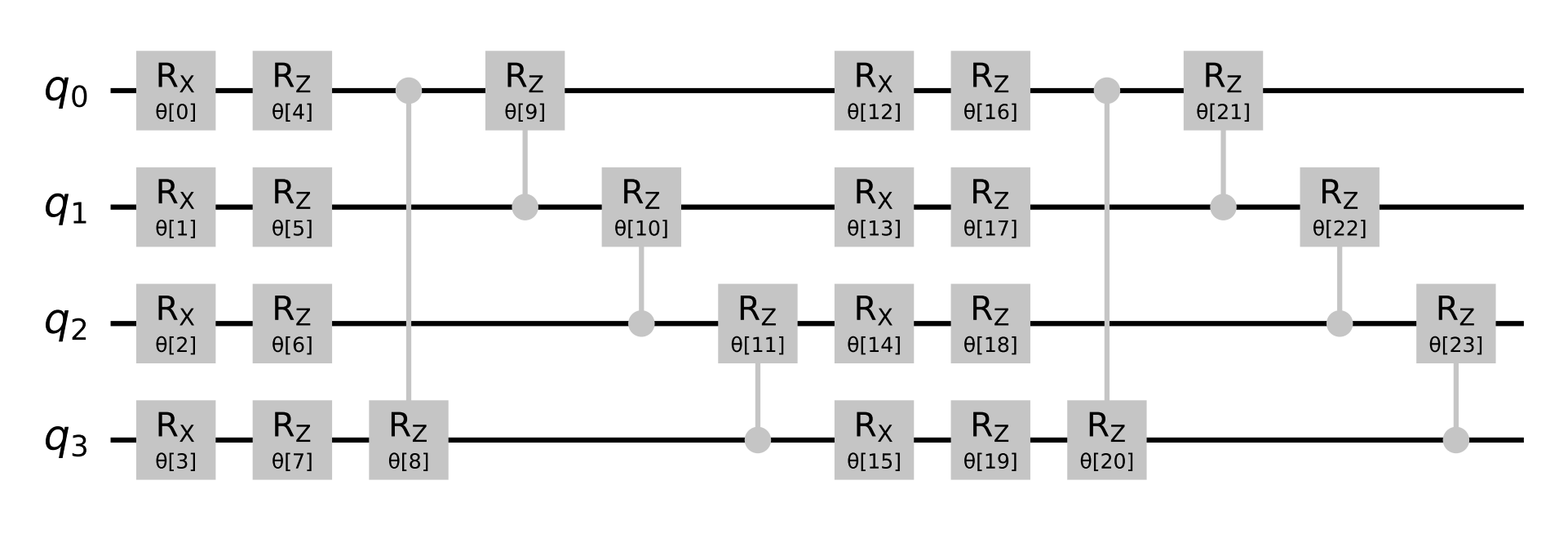}%
        \label{fig:18}%
    }\hfill  
    \subfloat[$C19$]{%
        \includegraphics[width=0.32\textwidth]{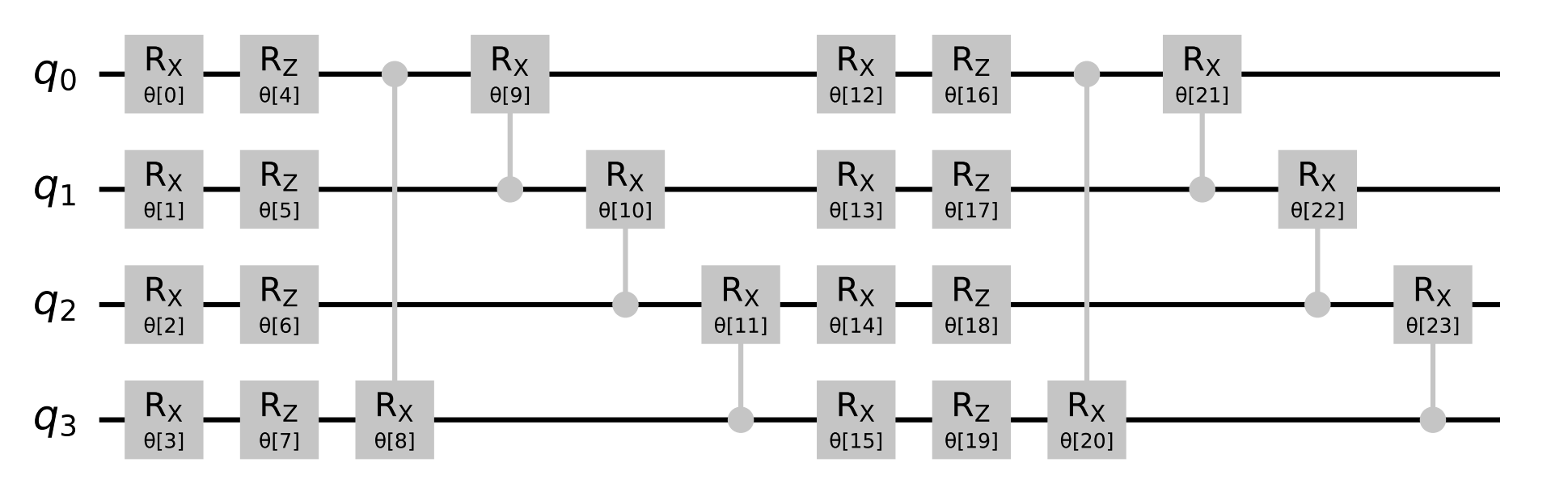}%
        \label{fig:19}%
    }\\  
    
    \caption{Variational quantum circuits.}  
    \label{fig:all}  
\end{figure}  

\bibliographystyle{ieeetr}
\bibliography{ref}
\end{document}